\documentclass[aps,twocolumn,nofootinbib,preprintnumbers,superscriptaddress]{revtex4-2}

\usepackage{color}
\usepackage{bbm}
\usepackage{tikz}

\newcommand{\nb}[1]{\color{blue}}

\newcommand{\hl}[1]{\color{black}}

\usepackage[
pagebackref=false,
colorlinks=true,
linkcolor=blue,
urlcolor=blue,
filecolor=black,
citecolor=red,
pdfstartview=FitV,
pdftitle={},
pdfauthor={},
pdfsubject={},
pdfkeywords={},
pdfpagemode=None,
bookmarksopen=true
]{hyperref}

\usepackage[normalem]{ulem}
\usepackage{amsmath, amssymb, bm}
\usepackage{enumerate}
\usepackage{amsfonts}
\usepackage{epsfig}
\usepackage{mathbbol}
\usepackage{mathrsfs}
\usepackage{braket}

\usepackage[utf8]{inputenc}
\DeclareUnicodeCharacter{0304}{\ensuremath{\bar{}}}

\begin{document}

\title{
\textbf{
Energy-twisted boundary condition
  and response  in one-dimensional quantum many-body systems
}
}

\author{Ryota Nakai}
\thanks{The first two authors contributed equally to the work.}
\affiliation{Department of Physics, Kyushu University, Fukuoka, 819-0395, Japan}

\author{Taozhi Guo}
\thanks{The first two authors contributed equally to the work.}
\affiliation{Department of Physics, Princeton University, Princeton, New Jersey, 08540, USA}

\author{Shinsei Ryu}
\affiliation{Department of Physics, Princeton University, Princeton, New Jersey, 08540, USA}

\date{\today}

\begin{abstract}
Thermal transport in condensed matter systems is traditionally formulated as a response to a background gravitational field. In this work, we seek a twisted-boundary-condition formalism for thermal transport in analogy to the $U(1)$ twisted boundary condition for electrical transport. Specifically, using the transfer matrix formalism, we introduce what we call the energy-twisted boundary condition, and study the response of the system to the boundary condition. As specific examples, we obtain the thermal Meissner stiffness of (1+1)-dimensional CFT, the Ising model, and disordered fermion models. We also identify the boost deformation of integrable systems as a bulk counterpart of the energy-twisted boundary condition. We show that the boost deformation of the free fermion chain can be solved explicitly by solving the inviscid Burgers equation. We also discuss the boost deformation of the XXZ model, 
and its nonlinear thermal 
Drude weights, 
by studying the boost-deformed Bethe ansatz equations. 
\end{abstract}

\maketitle

\section{Introduction}

Condensed matter systems are characterized by their responses to various background fields.
For example, electrical conductivity is a (linear) response to an applied electric field. 
More formally, the system can be gauged or coupled to arbitrary background
$U(1)$ gauge field, and one can study the response of the system.

The electrical response is far from the complete characterization of the system.
In particular, for charge-neutral systems or particle number non-conserving systems, we need to seek other responses.
For example, thermal transport can be well-defined and investigated for generic systems. 
Luttinger \cite{luttinger64} identified the gravitational field (gravitoelectric
field) as a proper static background field to formulate the linear response for thermal transport. 
(This is based on the Tolman-Ehrenfest effect, which is similar to the Unruh
effect.)
This formalism allows us to study thermal transport in much the same way as electrical transport. 

In this paper, we will further pursue parallelism between thermal and electrical response.
In particular, we seek an \textcolor{black}{analogue} of the twisted-boundary-condition formalism a la Kohn and Thouless
\cite{1964PhRv..133..171K, 1972JPhC....5..807E,PhysRevLett.39.1167}.
In this approach, the system's sensitivity to the twisted boundary condition
-- the boundary condition twisted by the particle number conserving
$U(1)$ phase rotation --
is related to the electrical transport. 
In this paper, we will discuss the boundary condition
twisted by energy, which we call
the energy-twisted boundary condition. 
Following the analogy, the sensitivity of the system
to the energy-twisted boundary condition is expected to
capture the system's transport properties.

For the case of electrical transport,
twisting the boundary condition by $U(1)$ phase
is gauge equivalent to 
introducing bulk background $U(1)$ gauge field. 
In particular, the bulk $U(1)$ gauge field can be completely
uniform (homogeneous).
Similarly, 
in relativistic theories,
the energy-twisted boundary condition can be
thought of as a change in the background metric
-- we introduce the background graviphoton field 
\cite{PhysRevB.95.165405, 2015arXiv151202607G}.
This is equivalent to put the system in an accelerated frame.
However, our formalism, the energy-twisted boundary condition,
can be applied to any lattice quantum many-body systems, as far as
energy is conserved
-- we can ``accelerate'' or ``boost''
lattice quantum many-body systems
by using the energy-twisted boundary condition.

While the equivalence between
the energy-twisted boundary condition and
the bulk background metric
may not hold for lattice quantum many-body systems in general,
we will discuss an analogue of the bulk formulation
for the case of integrable lattice quantum many-body systems. 
Concretely, we will discuss the so-called boost deformation
for integrable lattice quantum many-body systems.

In this paper, we will be mostly interested in (1+1)D
systems, defined on a spatial circle (ring).
The twisted boundary condition,
twisted either by $U(1)$ or by energy,
can be thought of as arising
from magnetic or gravitomagnetic flux
threading through the ring. 
For the case of $U(1)$,
this is the setting 
where we can discuss persistent electrical current
\cite{1983PhLA...96..365B},
related to the Aharonov-Bohm effect.
With energy-twisted boundary condition,
we can also discuss a gravitational analogue of persistent current.
Just like the persistent current is based on
the Aharonov-Bohm effect,
the thermal/gravitational analogue
can be thought of as related to the Sagnac effect
\cite{Rizzi2003-ka}.
The Aharonov-Bohm effect
and the persistent current is
periodic in the unit of flux quantum.
When the threaded flux is an integer multiple of
the flux quantum,
the Hamiltonian is equivalent to 
the Hamiltonian without magnetic flux, 
as one can find a large gauge (unitary) transformation
which brings one into the other.
While it is rarely discussed,
there is a similar periodicity
for the Sagnac effect, and for the
gravitational persistent current.
It is related to the large diffeomorphism
(modular transformation) of the spacetime torus.

The rest of the paper is organized as follows.
In Sec.\ \ref{Boost boundary condition},
we first recall the twisted boundary condition by
$U(1)$ phase and its relation to the Drude weight and
Meissner stiffness.
Subsequently, we consider the generalization,
the boundary condition twisted by time-translation symmetry. 
We then introduce the thermal version of the Drude weight and
Meissner stiffness.
The precise prescription for the energy-twisted boundary condition
is discussed by using the tensor network representation
of the transfer matrices.
In addition,
one can formulate the bulk perspective
using the so-called boost deformation in integrable systems. 
In Sec.\ \ref{Boost boundary condition and thermal Meissner stiffness},
we present the calculation of the Meissner stiffness
for (1+1)D CFT and for the transverse-field Ising model. 
In Appendix \ref{Transfer matrix method for free fermion models},
we also present the calculation of the Meissner stiffness
for (1+1)D disordered free fermion models
by using the transfer matrix method.
In Sec.\ \ref{Boost deformation in integrable systems},
we take a closer look at the integrable boost deformation,
by first focusing on the free fermion chain. 
We will show that the boost deformation can be
solved in terms of the inviscid Burgers equation.
We also study the boost deformation
for the XXZ model,
and its thermal response,
in particular, 
the nonlinear 
thermal Drude weights.
Finally, we conclude in Sec.\ \ref{Conclusion}.

\section{Energy-twisted boundary condition}
\label{Boost boundary condition}


\subsection{$U(1)$ twisted boundary condition,
  persistent current, Drude weight and Meissner stiffness}
\label{U(1) twisted boundary condition, persistent current, Drude weight and Meissner stiffness}

Any symmetry in quantum field theories can be twisted.
This is so in particular for unitary on-site symmetries.
By twisting, we here mean twisting boundary conditions
by symmetries.
(One can also introduce symmetry twist defects,
which are closely related.)
Of interest to us in this paper is twisting by time translation symmetry (energy).
Before discussing twisting by energy, let us start, as a warm-up, 
with a more familiar example of twisting by continuous $U(1)$ symmetry. 

To be specific,
let us consider a lattice fermion system
defined on a finite one-dimensional lattice
of length $L$
with the periodic boundary condition (PBC).
I.e., the system is defined on a spatial ring or circle.
(The following discussion can easily be extended to systems defined on a
$d$-dimensional spatial torus.)
We use
${\psi}_i(x)$ to denote a fermion annihilation operator 
located at a site $x$,
$i$ represents some internal degrees of freedom
within unit cell (spin, orbitals, etc.).
\textcolor{black}{For general systems,}
the boundary condition can be twisted,
i.e., 
we  can consider a twisting boundary condition,
${\psi}_i (x+ L) = e^{i\phi} {\psi}_i(x)$,
where $\phi$ is a twisting phase
\textcolor{black}{(notice, however, that we have systems with conserved particle number in mind in the following to discuss conduction properties.)}
By using the generator of $U(1)$, 
i.e., 
the total charge (total fermion number operator), 
${Q}=\sum_{x}\sum_i {\psi}^{\dag}_i(x) {\psi}^{\ }_i(x)$,
this boundary condition can be written as
\begin{align}
  \label{twisted by U1}
 {\psi}_i (x+ L) = 
 \mathcal{G}^{\ }_{\phi}\,
 {\psi}_i(x)\,
  \mathcal{G}^{-1}_{\phi},
  \quad
  \mathcal{G}_{\phi}= e^{ i \phi {Q}}.
\end{align}
\textcolor{black}{As is well known}, such twisting boundary condition 
can be realized by the Aharanov-Bohm effect, i.e., by putting
magnetic flux through a non-trivial cycle of the circle. 
Such magnetic flux may be introduced by a constant background 
gauge potential, e.g., 
$A(x) = \phi /L$.
This gauge potential enters into the hopping elements: 
$
{\psi}^{\dag}_j(x+1)  e^{i \phi/L} {\psi}^{\ }_i (x) 
+ h.c.
$
By a gauge transformation
$
{\psi}_i(x) \to e^{i\phi x/L} {\psi}^{\ }_i(x), 
$
one can remove the background vector potential,
$
{\psi}^{\dag}_j(x+1)  e^{i \phi/L} {\psi}^{\ }_i(x) 
\to
{\psi}^{\dag}_j(x+1) {\psi}^{\ }_i(x) 
$,
except at the boundary of the system: 
$
{\psi}^{\dag}_j(1) e^{i \phi/L} {\psi}^{\ }_i(L) 
\to
{\psi}^{\dag}_j(1) e^{i \phi} {\psi}^{\ }_i(L) 
=
{\psi}^{\dag}_j(1)
\mathcal{G}^{\ }_{\phi}
{\psi}_i(L)
\mathcal{G}^{-1}_{\phi}. 
$
After this gauge transformation,
only the link connecting 
the ends at 
$x=1$
and 
$x=L$
has a phase factor $e^{i\phi}$. 

The twisted boundary condition \eqref{twisted by U1}
can immediately be generalized to
any unitary on-site symmetries
by simply replacing $\mathcal{G}_{\phi}$ by
the unitary operator implementing the symmetry.
It can also be generalized to
non-on site symmetries
\cite{2018PhRvB..98c5151S},
and to 
antiunitary symmetries (time-reversal symmetry)
\cite{2017PhRvL.118u6402S, 2018PhRvB..98c5151S}.
These twisting are useful, e.g., to detect symmetry-protected topological phases.



With the twisted boundary condition,
we can now discuss the system's response
to the $U(1)$ twist, and associated
quantities that measure the response
\cite{1964PhRv..133..171K,trivedi88,PhysRevB.47.7995,giamarchi95,shastry06,Resta_2018}.
(Here, we follow the notation of \cite{shastry06}.)
First,
when the boundary condition is twisted by a $U(1)$ phase,
${\psi}_i(x+L)=e^{i\phi}{\psi}_i(x)$, inversion symmetry is broken and a finite
electric current,
the persistent current,
\begin{align}
 J=L\frac{dF}{d\phi}=L\sum_{n}\frac{e^{-\beta E_n}}{Z}\frac{dE_n}{d\phi},
\end{align}
flows in the ground state, where $E_n(\phi)$ is the many-body eigenenergy as a
function of the twisted $U(1)$ phase,
$Z=\sum_ne^{-\beta E_n}$ is the partition function, and $F=-\beta^{-1}\ln Z$ is the free energy.
By taking the second derivative with respect to the
$U(1)$ phase $\phi$,
we can measure the stiffness of
a system against the $U(1)$ twist. 
There are two similar but different quantities, the Drude weight (charge
stiffness) $\bar{D}$ and the Meissner stiffness $D$.
They are defined, respectively, by
\begin{align}
 &\bar{D}=\frac{L}{2}\sum_{n}\frac{e^{-\beta E_n}}{Z}\frac{d^2 E_n}{d\phi^2}\bigg|_{\phi=0},\\
 &D=\frac{L}{2}\frac{d^2 F}{d\phi^2}\bigg|_{\phi=0}.
\end{align}
In transport theory of free fermions, the Drude weight describes the singular part of the ac
electric conductivity
$\sigma(\omega)$
at zero frequency $\omega=0$,
\begin{align}
 \text{Re}\,\sigma(\omega)
 =
 2\pi \bar{D}\delta(\omega)+\sigma_\text{reg}(\omega).
\end{align}
On the other hand,
the Meissner stiffness measures the superfluid density and describes
the boundary-$U(1)$-phase dependent part of the ac conductivity as
\begin{align}
 \sigma(\omega)
 =
 \frac{2i D}{\omega+i\delta}
 +
 \sigma_\text{KG}(\omega).
 \label{eq:complexacelectricconductivity}
\end{align}
The second term of the right-hand side is the Kubo-Greenwood formula of the ac conductivity

In the limit of $L\to\infty$ and then $T\to 0$, the Drude weight is a measure of metallicity \cite{1964PhRv..133..171K}, and the Meissner stiffness is that of superconductivity \cite{PhysRevB.47.7995}, that is, $D=\bar{D}=0$ in insulators, $D=0,\bar{D}\neq 0$ in metals, and $D=\bar{D}\neq 0$ in superconductors. The coincidence of the two stiffnesses occurs when the energy gap is present \cite{PhysRevB.47.7995}.
In the limit of $L\to\infty$ but at a finite temperature, $\bar{D}$ is a measure of ballistic conduction or integrability \cite{PhysRevLett.74.972,zotos97,Fujimoto_1998,PhysRevB.77.245131}, while $D=0$ in one dimension.
As for a finite-size system, there is typically an energy gap above the ground state. Thus, at $T\to 0$, the Drude weight and the Meissner stiffness coincide provided 
there is no ground state degeneracy \cite{giamarchi95}.



\subsection{Energy-twisted boundary condition}

We shall now generalize the above line of thinking to time translation symmetry.
Following \eqref{twisted by U1},
we are interested in the
``energy-twisted'' boundary condition,
\begin{align}
  \label{energy twist bc}
  {\psi}_i(x+L)= e^{ a {H}}
  {\psi}_i(x) e^{-a  {H}},
\end{align}
where ${H}$ is the Hamiltonian,
and $a$ is a parameter.


To give a precise meaning of \eqref{energy twist bc},
we can switch to the imaginary-time (Euclidean) path-integral language, 
where the energy-twisted boundary condition can be
introduced, in term of the spacetime field, as
\begin{align}
  \label{time shifted}
  \psi_i(x+L, \tau)= \psi_i(x, \tau+a),
\end{align}
where $\tau$ is the imaginary time.
We should note that the imaginary time is periodic,
with the periodicity given by the inverse temperature $\beta$,
$\tau \equiv \tau + \beta$. 
Accordingly,
while not apparent in \eqref{energy twist bc},
there is a periodicity in the twist parameter $a$
with the periodicity, $a \equiv a + \beta$,
much the same way as the $U(1)$ twisted boundary condition
is periodic  
with periodicity given by the flux quantum,
$2\pi =
2\pi \hbar c/|q|$ where
$\hbar = c =1$ and we choose
the charge of the matter field to be one,
$|q|=1$.
We will call the boundary condition
of type \eqref{energy twist bc} or \eqref{time shifted}
as energy-twisted boundary condition.
We will also work with the rescaled version of $a$, 
\begin{align}
  \kappa = \frac{a}{L},
\end{align}
in terms of which the periodicity condition
is given by $\kappa \equiv \kappa + \beta/L$.

\begin{figure}[tbp]
  \centering
  \includegraphics[scale=0.75]{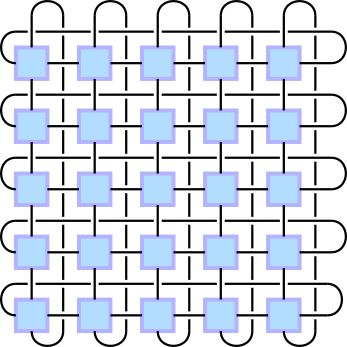}
  \hspace{0.5cm}
  \includegraphics[scale=0.75]{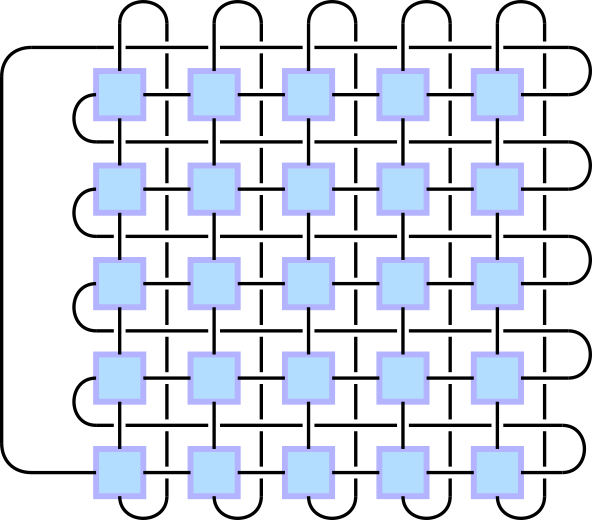}
  \caption{
    \label{TN}
    The tensor network
    representation of the
    untwisted (left) and twisted (right)
    partition functions.
  }
\end{figure}

It is also useful to consider discretized imaginary time
and the transfer matrix,
as commonly done
in lattice quantum many-body systems. 
The energy-twisted boundary condition can then be conveniently introduced
when we have a matrix-product-operator representation of
the (column-to-column) transfer matrix.
If we discretize the imaginary-time direction
into $M$ lattice sites,
$\beta = \Delta \tau \times M$, 
the partition function can be written 
in terms of the row-to-row transfer matrix
$
V \sim e^{ -\Delta \tau H}
$
as
$
  Z = \mathrm{Tr}
  \left[
  V^{M}
  \right]
$.
%
%
%
When the system's transfer
matrix is represented in terms of a matrix product operator,
the partition function on the torus
is then given in terms of a tensor-network,
as depicted in Fig.\ \ref{TN}.
The partition function can be 
alternatively
written in terms of the column-to-column transfer matrix $W$,
\begin{align}
  Z = \mathrm{Tr}
  \left[
  V^{M}
  \right]
  =
  \mathrm{Tr}\left[
  W^{L} 
  \right].
\end{align}

Now, we distort this spacetime lattice,
and consider the partition function on the twisted torus
(Fig.\ \ref{TN} right).
This can be achieved by ``reconnecting'' the relevant links,
located between $x=L$ and $x=1$.
This reconnection implements
a discrete version
of the energy-twisted boundary condition.
We note that twisted {\it spatial} tori
have been discussed in the context of topological order \cite{PhysRevB.88.195412,you15} and
Lieb-Schultz-Mattis type theorems
\cite{2021PhRvL.126u7201Y, 2021PhRvB.104g5146A}.

Viewing the horizontal direction as a fictitious time direction,
this may be viewed as an insertion of an operator in the column-to-column picture
\begin{align}
  Z_{{\it twist}}(a) = \mathrm{Tr} \left[W^{L} S^{a} \right],
\end{align}
where $S$ is the unit shift operator in time direction,
that shifts the temporal coordinate by $\Delta \tau$.
The twist parameter $a$ here is an integer.
We note when $a = M$, $S^{M}=\mathbb{1}$,
and hence the twisted partition function is periodic in
$a$, $Z_{{\it twist}}(a+M)= Z_{{\it twist}}(a)$.

\subsection{Energy-twisted boundary condition and deformation in integrable systems}

\label{Boost boundary condition and deformation in integrable systems}

While the above prescription to introduce
energy-twisted boundary condition is generic,
we now turn our attention to integrable
lattice systems and quantum field theories in (1+1) dimensions.
There, the energy-twisted boundary condition can be implemented
without breaking their integrability. 
Integrability of these models also allows us to
consider their boost deformations --
{\it bulk} deformations of the models without breaking
integrability
\cite{2009JPhA...42B5205B}.
Boost deformation is to energy-twisted boundary condition  
what bulk $U(1)$ gauge field is to
$U(1)$ twisted boundary condition. 
Namely, boost deformations provide a bulk background
``gauge field'' corresponding to the energy-twisted
boundary condition.

Let us now briefly
review the boost deformation in
integrable
(1+1)D lattice quantum many-body systems,
by first using the set of conserved charges,
and then
by
using the coordinate Bethe ansatz.
The latter description makes its connection to
the energy-twisted boundary condition clear, while in the former we have
a bulk description in terms of a deformed Hamiltonian.


We recall that
integrable spin chains
come with an infinite tower of
commuting charges $\{Q_r\}$,
$[Q_r, Q_s]=0$ ($r,s=2,3,\ldots$),
the existence of which is the manifestation of the integrability.
Among the conserved charges is the Hamiltonian of the spin chain,
$H=Q_2$.
Ref.\ \cite{2009JPhA...42B5205B}
introduced one parameter
deformations of generic integrable quantum spin chains.
Starting from the infinite tower of
commuting charges $\{Q_r\}$ of the original
short-range spin chain,
the scheme introduced in
Ref.\ \cite{2009JPhA...42B5205B}
continuously deforms 
the conserved charges $\{Q_r\} \to \{Q_r(\lambda)\}$
where $\lambda$ is the deformation parameter.
Under such deformation, the integrability is maintained,
i.e.,
$[Q_r(\lambda), Q_s(\lambda)]=0$, 
but the deformed charges are longer-ranged.
One of the examples of the deformations
is the so-called $T\bar{T}$ deformation
\cite{2004hep.th....1146Z}.
Of our interest here is the boost deformation,
which is defined, for the second conserved charge (the Hamiltonian),
by
\begin{align}
  \frac{d Q_r(\lambda)}{d\lambda}
  =
  i
  [
  \mathcal{B}[Q_2(\lambda)], Q_r(\lambda)
  ].
 \label{eq:boostdeformation}
\end{align}
Here, $\mathcal{B}[Q_2(\lambda)]$ is the boost operator
for the charge $Q_2$ and defined by
\begin{equation}
	\mathcal{B}[Q_2] = \sum_x\, x\, q_2(x),
\end{equation}
where $q_2(x)$ is the density of $Q_2$, $Q_2 = \sum_x q_2(x)$.
The boost-deformed Hamiltonian $Q_2(\lambda)= H(\lambda)$
is
an analogue of the Hamiltonian $H(\phi)$
in the presence of background $U(1)$ gauge
field $A(x)= \phi /L$
discussed in
Sec.\ \ref{U(1) twisted boundary condition, persistent current, Drude weight and Meissner stiffness}.
As will be seen in Eq.\ \eqref{twisted k quantization}, the parameter $\lambda$
can be identified with the
parameter $a,\kappa$
introduced in Sec.\
\ref{Boost boundary condition}
as
\begin{align}
  i\lambda =  \kappa,
\end{align}
i.e., an analytic continuation of
$\kappa$.
We note that the flow equation
\eqref{eq:boostdeformation}
for real $\lambda$
keeps the conserved charges hermitian,
while the operator twisting  
the boundary condition in \eqref{energy twist bc}
is non-unitary when $a$ and $\kappa$ are real.

%


In the above
the boost deformation is conveniently described
for infinite systems. 
It is however possible to discuss integrability and the deformation
for finite chains. 
There, we need to worry about the compatibility
between the long-range nature of the deformed conserved
charges, and the finite size of the system with a boundary condition.
As long as the range of a conserved charge of interest does not exceed the
length of the chain $L$,
one can formulate the Bethe ansatz equations,
and expect
that they give the correct spectrum
for this particular charge.
The Bethe ansatz equations we use here are asymptotic ones,
valid for large enough $L$.


Let us consider, as an example, 
the $S=1/2$ XXZ spin chain 
\begin{align}
\label{XXZ Ham}
   H =
   J
 \sum_{x=1}^L 
 \left(S_x^x S_{x+1}^x 
 + S_x^y S_{x+1}^y 
 + \Delta S_x^z S_{x+1}^z
 \right) - \frac{LJ\Delta}{2}.
\end{align}
In the following,
we will assume $L$ to be even, 
and
$J>0$ and $-1<\Delta<1$.
We parameterize  the anisotropy $\Delta$ as
$
\Delta = \cos \gamma
$.
The coordinate Bethe ansatz
for a state containing $N$ ``particles''
with (quasi) momenta $p_1, \ldots, p_N$
is given by
\begin{align}
  \left|\boldsymbol{v}_{N}\right\rangle
  &=\sum_{x_{1}<x_{2}<\cdots<x_{N}} \sum_{\sigma \in S_{N}}
    \prod_{j>k} f\left(v_{\sigma_{j}}-v_{\sigma_{k}}\right)
  \nonumber \\
  &\quad
    \times 
    \prod_{j=1}^{N} e^{i p_{\sigma j} x_{j}}
S^-_{x_1} \cdots S^{-}_{x_N} \ket{\uparrow \cdots \uparrow},
\label{BA state}
\end{align}
where $S_N$ is the symmetric group of degree $N$ and $S_{x_j}^-=S_{x_j}^x-iS_{x_j}^y$.
Here,
we introduce the rapidity variable
$v_j$,
\begin{equation}
e^{ip_j} =
e^{ i p(v_j)}
=  \frac{\sinh{\frac{\gamma}{2}(v_j+i)}}{\sinh{\frac{\gamma}{2}(v_j-i)}},
\end{equation}
and $f(v)$ is related to the $S$-matrix 
and the phase shift,
\begin{align}
S(v)
=
e^{ i \delta (v)}
   = 
   -
   \frac{ 
   \sinh \frac{\gamma}{2} (v + 2i)
   }{
   \sinh \frac{\gamma}{2} (v - 2i)
   },
\end{align}
by $S(v)=f(v)/f(-v)$.
The energy 
for the state \eqref{BA state}
is
given by
\begin{align}
E &= 
\sum_{j=1}^N h(v_j)
=
\sum_{j=1}^{N}\frac{2J\sin^2{\gamma}}
{\cos{\gamma}-\cosh{(\gamma v_j)}}.
\end{align}
Requiring PBC, 
we obtain the Bethe ansatz 
equations
\begin{equation}
  e^{i p(v_{j}) L}=
  \prod_{k (\neq j)} S\left(v_{j}-v_{k}\right), \quad
  j=1, \ldots, N,
\end{equation}
that determine the quasi momenta.

The boost deformation results in the change in momentum
$p(v_j) \to p_{\lambda}(v_j)$
\cite{2009JPhA...42B5205B}.
We can then consider 
the modified Bethe ansatz equations
\begin{equation}
  \label{twisted BAE}
  e^{ i p_{\lambda}(v_j)L}
  = \prod_{k(\neq j)} S(v_j-v_k).
\end{equation}
As mentioned above,
these Bethe ansatz equations are
asymptotic ones,
valid for large enough $L$.
In Ref.\ \cite{2020ScPP....8...16P},
it was shown that,
in infinite volume,
the deformed momentum $p_{\lambda}$ 
depends linearly on $\lambda$,
\begin{align}
p_{\lambda}(v_j)= p_{\lambda=0}(v_j)+ \lambda h(v_j),
\end{align}
which is an input to the Bethe ansatz equations. 

\subsection{Thermal response}

In analogy to the $U(1)$ case, we expect that
the energy-twisted boundary condition and deformation (\ref{eq:boostdeformation}) is related to thermal transport.
The commutator of the Hamiltonian with the boost operator
(the right-hand side of \eqref{eq:boostdeformation}
when $r=2$) is the energy current operator, which is an integral of motion in integrable models.
The persistent heat current 
(an analogue of the persistent charge current)
flowing in the ground state of a boost-deformed Hamiltonian is thus
\begin{align}
  J^Q=-\left\langle \frac{dQ_2}{d\lambda}\right\rangle
  =-\sum_{n}\frac{e^{-\beta E_n}}{Z}\frac{dE_n}{d\lambda}
  =
  -\frac{dF}{d\lambda}.
\end{align}

From the linear response theory, we can define the thermal Drude weight \cite{shastry06}, which is the zero-frequency singularity part of the ac thermal conductivity
\begin{align}
 \text{Re}\,\kappa(\omega)
 =
 \frac{2\pi \bar{D}^Q}{T}\delta(\omega)+\kappa_\text{reg}(\omega),
\end{align}
and the thermal version of the Meissner stiffness \cite{shastry06},
which
is the contribution to the thermal conductivity
besides the Kubo-Greenwood part
\cite{luttinger64}
\begin{align}
 \kappa(\omega)
 =
 \frac{2i D^Q}{T(\omega+i\delta)}+\kappa_\text{KG}(\omega).
 \label{eq:acthermalconductivity}
\end{align}
Notice that the definition of these quantities is due to \cite{shastry06}, which may be different from other references by $T$ and a constant.
As expected, the thermal Drude weight and Meissner stiffness of free fermions are identified with the second derivatives of the energy and free energy, respectively, with respect to the boost-deformation parameter $\lambda$ as
\begin{align}
 &\bar{D}^Q=\frac{1}{2L}\sum_{n}\frac{e^{-\beta E_n}}{Z}\frac{d^2 E_n}{d\lambda^2}\bigg|_{\lambda=0},
 \label{eq:thermaldrudeweight_boostdeformation}\\
 &D^Q=\frac{1}{2L}\frac{d^2 F}{d\lambda^2}\bigg|_{\lambda=0}.
 \label{eq:thermalmeissnerstiffness_boostdeformation}
\end{align}
(see Appendix \ref{sec:thermaltransport_deformation}).

At a finite temperature, the thermal Meissner stiffness is zero unless superconducting \cite{shastry06}. 
The thermal Drude weight has been studied in 1d quantum systems in \cite{Kl_mper_2002,PhysRevLett.89.156603,PhysRevB.66.140406,PhysRevB.67.064410,PhysRevB.67.134426,Sakai_2003}. 
[Specifically, see \eqref{eq:thermaldrudeweight_hight_1dsystems}.]

\section{Energy-twisted boundary condition and thermal Meissner stiffness}
\label{Boost boundary condition and thermal Meissner stiffness}

In this section, we consider the energy-twisted
boundary condition
in (1+1)D CFT and lattice many-body systems, and
calculate the thermal Meissner stiffness.

\subsection{(1+1)D CFT}
\label{sec:cft}

Let us start with a simple example, 
the (1+1)D chiral Dirac fermion theory, 
\begin{align}
  {H} =
  \int^L_0 dx\,
  {\psi}^{\dag}
  \mathcal{H}
  {\psi},
  \quad
  \mathcal{H} = - i v \partial_x,
\end{align}
where
$\psi(x)$ is a complex fermion field operator,
and 
$v$ is the Fermi velocity.
The single-particle eigen functions
are given by
$
  f_p (x) = e^{i p x}/\sqrt{L}
$
with the single-particle energy $\varepsilon(p) = v p$,
$\mathcal{H} f_p(x) = \varepsilon(p) f_p(x)$.
Requiring the regular (unboosted) PBC
leads to the quantization of $p$,
$p= 2\pi/L \times {\it integer}$.
The energy-twisted boundary condition can be imposed
by requiring
\begin{align}
  f_p(x+L) = e^{ipL} f_p(x)= e^{ 
  \kappa \varepsilon(p) L} f_p(x),
\end{align}
where $\kappa$ is the twist parameter.
Thus, $p$ is quantized as 
\begin{align}
  &
  \label{twisted k quantization}
  (p + i\kappa \varepsilon(p))L = 2\pi  n,
  \quad
n \in \mathbb{Z},
    \nonumber \\
  &
  \Rightarrow 
  p=
  \frac{1}{1 +  i v \kappa} \frac{2\pi n}{L}.
\end{align}
This equation should be compared with \eqref{twisted BAE}
with $i\lambda = \kappa$.

\begin{figure}
 \centering
 \includegraphics[width=45mm]{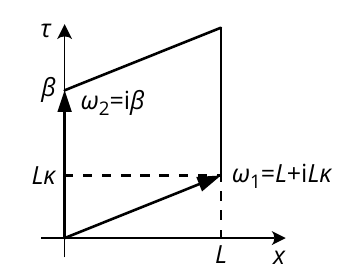}
 \caption{The spacetime torus with the energy-twisted boundary condition.
   (Here, we set the velocity to be one, $v=1$, for simplicity.)
 \label{fig:torus}}
\end{figure}

We now consider the partition function
in the presence of energy-twisted boundary condition. 
In relativistic systems
this can be incorporated by
introducing graviphoton field
in the background metric
(Appendix \ref{Graviphoton field on 2-torus}).
For \textcolor{black}{2-torus}, the twist can be incorporated
by modifying the modulus.
For the untwisted case,
the partition function is
\begin{align}
  Z = \mathrm{Tr}\, e^{ - \beta H}
  =
  \mathrm{Tr}\, q^{L_0-c/24},
\end{align}
where $L_0$ is the Virasoro generator, 
$c=1$ is the central charge,
and $q$ is given by
\begin{align}
  q
  =
  e^{ 2\pi i \tau} =
  e^{ - \frac{2\pi v\beta}{L}}.
\end{align}
With \eqref{twisted k quantization},
the partition function in the presence of
the twist is given by
$Z=\mathrm{Tr}\, q^{L_0-c/24}$,
where $q$ is now given by
\begin{align}
  q= e^{2\pi i \tau}=e^{- \frac{ 2\pi v \beta}{L} \frac{1}{1+i v \kappa} }.
\end{align}
Namely, the modulus
changes
from the untwisted to twisted case as
\begin{align}
\label{eq:energytwistedmodularparameter}
   \tau =  \frac{ iv \beta }{L} 
  \to 
  \frac{ i v \beta}{L} \frac{1}{1+i v \kappa}.
\end{align}
Recall that
the modulus
is the ratio of two periodicities $\omega_1$ and $\omega_2$
on the complex plane,
$\tau=\omega_2/\omega_1$.
After the twist, $\omega_1$ is changed as $L\to L+iL v\kappa$, while
$\omega_2=iv\beta$ remains unchanged (Fig.~\ref{fig:torus}).
When $\kappa= \beta/L$,
\eqref{eq:energytwistedmodularparameter}
is nothing but the modular transformation $TST$,
\begin{align}
   \tau \to \frac{\tau}{1+\tau},
\end{align}
where $T:\tau\to \tau+1$ and $S:\tau\to -1/\tau$ are the generators of the modular group $SL(2,\mathbb{Z})/\mathbb{Z}_2$: $\tau\to(a\tau+b)/(c\tau+d)\,(a,b,c,d\in\mathbb{Z})$. 
The modular transformation 
leaves the spacetime torus unchanged
(it acts as a large diffeomorphism),
and hence the spacetime at $\kappa$
and $\kappa + \beta/L$ are equivalent.

%
%
%

Let us now consider the energy-twisted boundary condition
in a generic (1+1)D CFT
using the formalism in Sec.\ \ref{Boost boundary condition}.
In Lorentz invariant theories,
row-to-row and column-to-column transfer matrices are 
essentially the same.
The row-to-row transfer matrix is given in
terms of the Hamiltonian $H$ as
$V= \exp (- H)$.
For a CFT placed on the spatial circle of circumference $L$,
$H$ is given 
in terms of the Virasoro generators $L_0$ and $\bar{L}_0$
and the central charge $c$ as
$
{H}=
(2\pi v/L)
(L_0 + \bar{L}_0 - c/12).
$
($v$ is the velocity of the excitations and plays the role
of the speed of light.)
The corresponding column-to-column transfer matrix is given by
$W = \exp (- \tilde{H})$
where
$
\tilde{H}=
(2\pi/v \beta)
(L_0 + \bar{L}_0 - c/12).
$
The partition function can be written in two different ways,
$Z(\beta,L)=
\mathrm{Tr}_{\mathcal{H}}\, e^{-\beta H}
=
\mathrm{Tr}_{\tilde{\mathcal{H}}}\, e^{-L \tilde{H} }
$,
where $\mathcal{H}$ and $\tilde{\mathcal{H}}$
are the CFT Hilbert space
on a ring of circumference $L$ and $\beta$, respectively. 
Introducing the moduli as
\begin{align}
  &
   \tau =   i v \beta/L,
  \quad
    \bar{\tau} =   - i v \beta/L,
  \nonumber \\
  &
    \tilde{\tau} = -1/\tau,
    \quad
    \bar{\tilde{\tau}} = -1/\bar{\tau},
\end{align}
the partition function can be written as
\begin{align}
  {Z}(\beta, L)
  &=
  \mathrm{Tr}_{\mathcal{H}}\,
  e^{ 2\pi i \tau (L_0-c/24)}
  e^{ -2\pi i \bar{\tau} (\bar{L}_0-c/24)}
    \nonumber \\
  &=
  \mathrm{Tr}_{
  \tilde{\mathcal{H}} }\,
  e^{ 2\pi i \tilde{\tau} (L_0-c/24)}
  e^{ -2\pi i \bar{\tilde{\tau}} (\bar{L}_0-c/24)}.
 \label{eq:cftpartitionfunction_modularinvariance}
\end{align}
To introduce the energy twist,
we modify the moduli as
\begin{align}
  &
  \tau = \frac{ iv \beta }{L} 
  \to 
  \frac{ i v \beta}{L} \frac{1}{1+i v \kappa},
  \nonumber \\
  &
  \tilde{\tau} =
  \frac{ iL }{v\beta} 
  \to 
    \frac{iL}{v\beta}
    (1+i v \kappa).
\end{align}
The energy-twisted partition function
is invariant under
$\tilde{\tau}\to \tilde{\tau}+n$
or
$L\kappa \to L\kappa+ n \beta$,
where $n$ is an integer.
%

\begin{figure}[t]
  \centering
\includegraphics[scale=0.8]{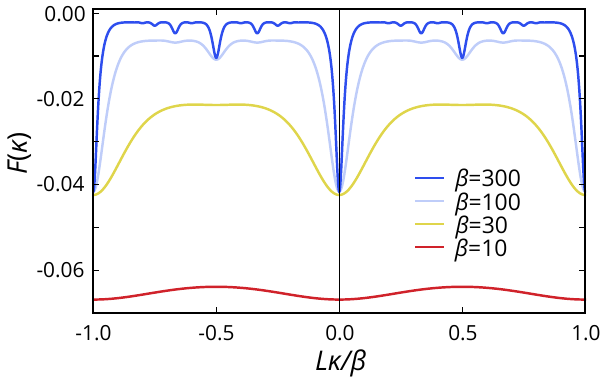}
  \caption{
    The variation of the free energy of
    the Ising CFT with energy twist.
 \label{fig:cfttwist}
  }
\end{figure}

The energy-twisted partition function
in the low-temperature limit,
$v\beta/L \to \infty$, can be evaluated as
\begin{align}
  \label{cft result}
  &
  Z(\beta, L) \sim
  e^{ 2\pi i \tau (h-\frac{c}{24}) }
  e^{ -2\pi i \bar{\tau} (h-\frac{c}{24}) },
  \nonumber \\
  &
  (-1/\beta)\ln Z
  \sim
  -\frac{ (c- 24 h) \pi v }{ 6L(1 + v^2 \kappa^2)}.
\end{align}
Here,
$h$ denotes the
(rescaled) ground state energy.
The thermal Meissner stiffness in the low-temperature limit converges to
\begin{align}
 D^Q(T=0)\sim -\frac{(c-24h)\pi v^3}{6L^2}.
 \label{eq:cft_thermalmeissnerstiffness}
\end{align}
This quantifies the variation of the ground state energy in response to the energy-twisted boundary condition.
According to Appendix \ref{sec:thermaltransport_deformation_thermalconductivity}, $D^Q(T=0)$ agrees with the same limit of the thermal Drude weight $\bar{D}^Q(T=0)$ due to the presence of a finite-size gap and the uniqueness of the ground state.
As an example, 
the twisted free energy 
of the Ising CFT is plotted in
Fig.\ \ref{fig:cfttwist}.
The behavior near $\kappa=0$ matches with \eqref{cft result} with $h=0$.

Between $L\kappa/\beta=0$ and $1$, the free energy at low temperature has smaller
Lorentzian peaks
\begin{align}
 F
 \simeq
 -\frac{(c-24h)\pi v}{6q^2L(1+v^2\delta \kappa^2)},
 \label{eq:cftfreenergypeaks_intermediate}
\end{align}
at $L\kappa/\beta=p/q+L\delta\kappa/\beta$,
where $p$ and $q$ are mutually coprime integers
and $\delta\kappa$ is a small deviation from $p/q$ (see Fig.\ \ref{fig:cfttwist}).
Specifically, a peak at $L\kappa/\beta=1/2$ is 1/4 the height at $\kappa=0$, peaks at $L\kappa/\beta=1/3$ and $2/3$ are 1/9 the height at $\kappa=0$, and so on. 
These peaks have the same origin as the fidelity after a quantum quench in CFT \cite{cardy14}. 
The quench dynamics at a time $t$ are traced by a modulus $\tau=v(i\beta+t)/L$, which is related to our twist by the $S$-modular transformation and interchanging $\beta$ and $L$. 
As a result, the fidelity has more peaks at higher temperature, while the energy-twisted free energy has more peaks at lower temperature.

Following \cite{cardy14},
the formula
\eqref{eq:cftfreenergypeaks_intermediate}
can be derived as follows.
A successive application of modular transformations $ST^{n_0}ST^{n_1}\cdots ST^{n_k}$ maps a modulus $\tau=q/p$ to $\tau=0$,
where $n_0,\cdots,n_k$ are integers appearing in the continued fraction of $p/q$ as
\begin{align}
 \frac{p}{q}
 =
 n_0
 -
 \cfrac{1}{n_1-\cfrac{1}{n_2-\cdots}}.
\end{align}
By the same modular transformation, a modulus $\tau=iv\beta/(L+iv(p/q)\beta+ivL\delta \kappa)$ is mapped to $\tau\simeq iq^2L(1+iv\delta \kappa)/v\beta$ when $v\beta/L\gg 1$, which relates the behavior around $L\kappa/\beta=p/q$ with that around $\kappa=0$.
Finally performing the $S$ transformation again, the free energy (\ref{eq:cftfreenergypeaks_intermediate}) is obtained, provided $v\beta/L\gg q^2$ and $L\delta \kappa\ll \beta$.

On the other hand, high-temperature ($v\beta/L\ll 1$) behavior can be addressed provided the modular invariance is present.
From (\ref{eq:cftpartitionfunction_modularinvariance}), we obtain $D^Q\sim 0$, which agrees with \cite{shastry06}.
Notice that high temperature in CFT indicates a temperature regime that is much higher than the energy-level spacing. At high temperature in CFT but, simultaneously, sufficiently lower than other energy scales, such as the band width or Ising coupling, the thermal Drude weight estimated from the heat current 
\textcolor{black}{has been reported \cite{PhysRevB.66.140406,PhysRevB.67.134426}, and is given by}
\begin{align}
 \bar{D}^Q=\frac{(c-24h)\pi vT^2}{6}.
 \label{eq:thermaldrudeweight_hight_1dsystems}
\end{align}

\subsection{The transverse-field Ising model}

While in the above we demonstrated the basic ideas
using (1+1)D CFT as an example, 
it is interesting to apply the idea to broader systems,
which do not have conformal symmetry nor Lorentz invariance.
Here, we consider the transverse-field Ising model
\begin{align}
 H
 =
 \sum_{i=1}^L
 \left(
 -J\sigma_i^x\sigma_{i+1}^x
 -
 h \sigma_i^z
 \right),
\end{align}
satisfying PBC ($\sigma_{L+1}=\sigma_1$).
The Ising coupling favors a ferromagnetically ordered phase ($J>h$), and the transverse field favors a disordered (paramagnetic) phase ($J<h$).
These phases are related to each other by an order-disorder duality transformation \cite{schultz64}.
The phase transition between them occurs at $h/J=1$ (the self-dual point), at which the low-energy properties are described by the Ising CFT \cite{francesco97}. 

\begin{figure}
 \centering
 \includegraphics[width=0.7\columnwidth]{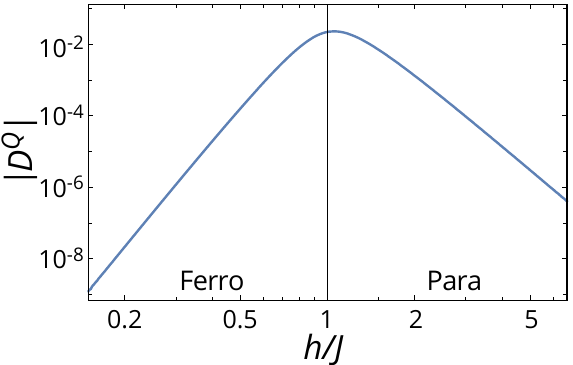}
 \includegraphics[width=0.96\columnwidth]{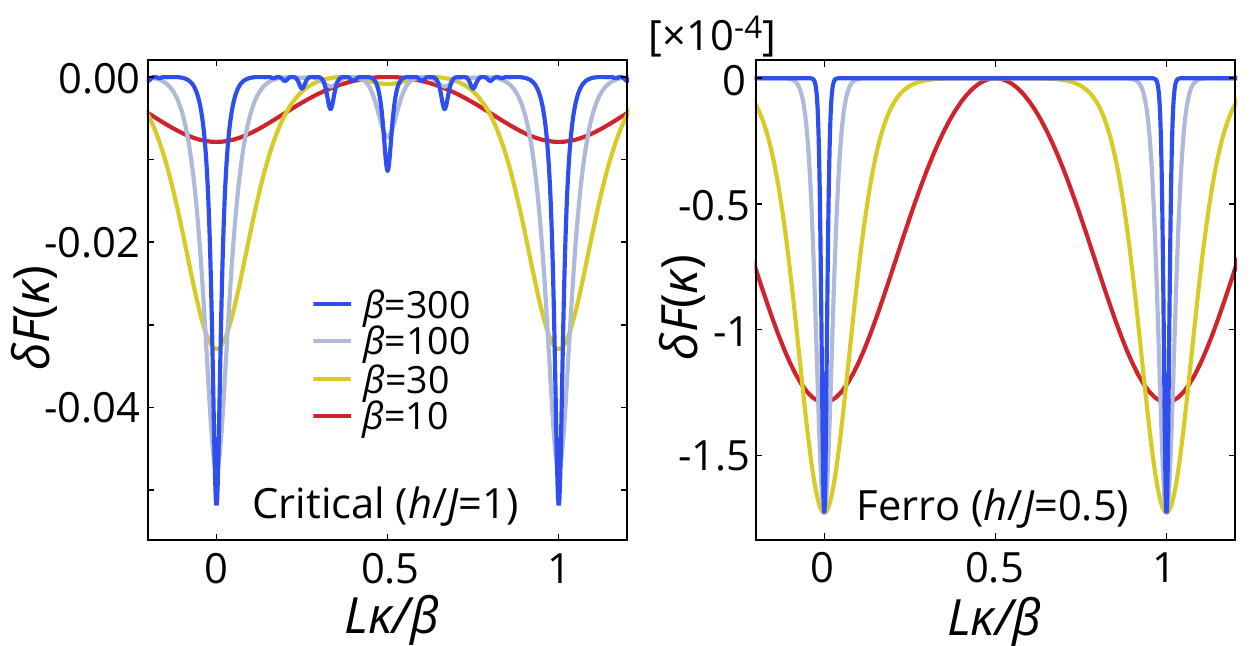}
 \caption{(Top) The thermal Meissner stiffness of the transverse-field Ising model ($L=10$ and $J=1$) at $\beta=100$.
 (Bottom) The variation of the free energy of the transverse Ising model at the critical point ($J=h=1$, \textcolor{black}{left}) and in a ferromagnetic phase ($J=1, h=0.5$, \textcolor{black}{right}) and the perimeter $L=10$ against the twist parameter $\kappa$ is shown for temperature $\beta=10,30,100$, and $300$ (from red to blue).
 \label{fig:xytwist}}
\end{figure}

We use the transfer matrix formalism introduced in
Sec.\ \ref{Boost boundary condition}
to calculate the response to the energy-twisted boundary condition.
Some details can be found in Appendix 
\ref{Lattice spin systems and transfer matrix formalism}.
The twisted free energy $F(\kappa)=-\beta^{-1}\ln Z(\kappa)$
is evaluated numerically for a ring of perimeter $L=10$.
Here, we fix the Ising coupling by $J=1$.
The free energy at the critical point ($h=J=1$) agrees with the CFT result (Fig.~\ref{fig:xytwist} bottom left).
The free energy has a period of $\kappa=\beta/L$ and the peaks of the free energy become clear as the temperature is lowered.
The free energy changes non-monotonically as a function of the twist parameter
$\kappa$, which is in stark contrast to a monotonically varying free energy of electrons under the $U(1)$ twist within a single quantum flux, exhibiting a saw-tooth shape.
\textcolor{black}{The free energy shifted by a suitable constant is plotted.}

Away from the critical point, we can see that the free energy variation decays rapidly due to the stiffness of the order (Fig.~\ref{fig:xytwist} bottom right).
In addition, the free energy peaks besides $L\kappa/\beta=$ (integer) fade out even at low temperature.
The peak height at these points is no longer related to that at the origin \textcolor{black}{as it is at the critical point}.
This would be a signature of the deviation of the theory from the Ising CFT.
The free energy profile obeys the duality of the model, that is, $h/J<1$ in the ferromagnetic phase and $h'/J'=(h/J)^{-1}>1$ in the paramagnetic phase have the same response against the twist.

In Appendix \ref{Transfer matrix method for free fermion models},
we consider yet another lattice model,
the 1d disordered free fermion model(s),
and discuss the thermal Meissner stiffness.

\section{Boost deformation in integrable systems}
\label{Boost deformation in integrable systems}

In this section, we discuss the boost deformation
in (1+1)D integrable lattice systems.
In particular, we first look at the free fermion model
in detail and show that the boost deformation
leads to the Burgers equation of the single-particle
dispersion.
We then turn to the XXZ model, and,
by using the boost-deformed Bethe ansatz equations,
calculate the ground state energy as a function of
the boost parameter, and the thermal Drude weight.

\subsection{The free fermion model}

\paragraph{The boost deformation and the inviscid Burgers equation}

We start from the (undeformed) tight-binding model
on a 1d lattice ($x\in\mathbb{Z}$), 
$
H = - \sum_x (c^{\dag}_x c_{x+1} + {\it h.c.})
$,
which defines
the initial condition
$Q_2(\lambda=0)=H$
of the boost deformation \eqref{eq:boostdeformation}.
It is straightforward to verify that
the Hamiltonian (the second charge) stays
quadratic during the boost deformation. 
Hence, we represent the Hamiltonian
and the corresponding boost operator as
\begin{align}
    \label{eq:thesecondconservedcharge}
  &Q_2(\lambda) =
  \sum_{x,z} t_z (\lambda) c^{\dag}_x c^{\ }_{x+z},
  \nonumber \\
  &\mathcal{B}[Q_2(\lambda)]
  =
  \sum_{x,z} (x+z/2) t_z (\lambda) c^{\dag}_x c^{\ }_{x+z},
\end{align}
where the set of
$\lambda$-dependent coefficients $t_z(\lambda)$
parameterize the boost-deformed Hamiltonian
with the initial condition
$t_z(\lambda=0) = - \delta_{1,z} - \delta_{-1,z}$.
In terms of
the coefficients $t_z(\lambda)$,
the flow equation of
the boost deformation \eqref{eq:boostdeformation},
reduces to
\begin{align}
  \frac{d t_z(\lambda)}{d\lambda} 
  =
  -\frac{iz}{2}
  \sum_{w} t_{w} (\lambda) t_{z-w}(\lambda).
 \label{eq:deformation_coefficientequation}
\end{align}

Starting from the nearest neighbor
tight-binding model $Q_2(\lambda=0)$,
the boost deformation \eqref{eq:boostdeformation}
(or the coupled ODE \eqref{eq:deformation_coefficientequation})
generates a longer-range hopping Hamiltonian
$Q_2(\lambda)$.
For a given $\lambda$, we consider 
a large enough chain of length $L$, 
and impose PBC.
By the Fourier transform
$
c_x =L^{-1/2} \sum_k e^{ i k x} \tilde{c}_k
$,
[$k=2\pi ({\it integer})/L$],
the Hamiltonian in momentum space is 
$H(\lambda) =
  \sum_{k} f(\lambda, k) \tilde{c}^{\dag}_k \tilde{c}^{\ }_{k},$
where the energy dispersion $f(\lambda, k)$
is given by the Fourier transform of $t_z(\lambda)$: 
\begin{align}
 f(\lambda, k):= \sum_w e^{i w k} t_w(\lambda).
\end{align}
%
%

From the perspective from the coupled ODE
\eqref{eq:deformation_coefficientequation},
the dispersion $f(\lambda,k)$ can be considered as
the ``generating function'' of
the coefficients $t_z(\lambda)$.
So far as the generating function
is differentiable with respect to $k$,
it obeys a PDE,
the inviscid Burgers equation 
\begin{align}
  \frac{\partial f}{\partial \lambda}
  +
  f \frac{\partial f}{\partial k}
  =0,
 \label{eq:burgersequation}
\end{align}
which can be derived from
\eqref{eq:deformation_coefficientequation}.

The inviscid Burgers equation has a formal solution derived by the method of characteristics \cite{Landau1987Fluid}.
The equi-energy contour in the $\lambda$-$k$ space is $k=\lambda f(\lambda=0,k_0)+k_0$ that emanates from a point $(\lambda,k)=(0,k_0)$.
This equation indicates how an initial state with a momentum $k_0$ and an eigenenergy $f(0,k_0)$ evolves by fixing the eigenenergy.
A state of $(k,f(\lambda,k))$ on the dispersion relation moves along the momentum direction at a speed of $f(\lambda,k)=f(0,k_0)$.
Thus the deformed dispersion relation is obtained by tilting the energy axis by $\arctan\lambda$.
The Hamiltonian can be deformed until the dispersion relation becomes singular, where the slope of the dispersion relation diverges.
Beyond this point, the generating function is no longer differentiable (the formation of the shock wave by the terminology of the hydrodynamics).

\begin{figure}
  \centering
  \includegraphics[width=70mm]{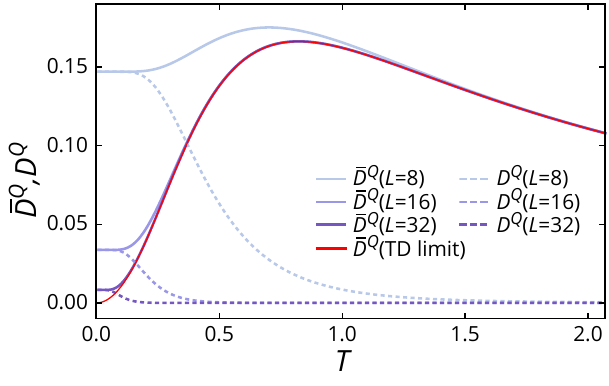}
  \includegraphics[width=70mm]{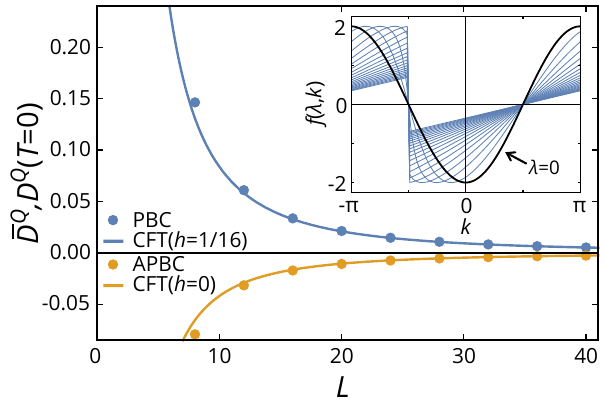}
  \caption{
  (Top) The thermal Drude weight and thermal Meissner stiffness 
  of the free fermion chain with $L=8,16,32$ (solid and dotted lines, respectively) and the thermal Drude weight in the thermodynamic limit (red line).
  (Bottom) The thermal Drude weight (thermal Meissner stiffness) at zero temperature is shown. 
  The inset is the evolution of the dispersion relation from $\lambda=0$ (black) to $\lambda=4$.
 \label{fig:deformedcosine}
  }
\end{figure}

When the generating function is not differentiable with respect to
$k$, it obeys an integro-differential equation
\begin{align}
 \frac{\partial}{\partial \lambda}
 \int dk\,
 e^{-iw k}
 f(\lambda,k)
 =
 \frac{1}{2}
  \int dk
 \frac{\partial e^{-iw k}}{\partial k}
 f(\lambda,k)^2.
 \label{eq:inviscidburgers_integral}
\end{align}
Solutions to (\ref{eq:inviscidburgers_integral}) are known as weak solutions to the inviscid Burgers equation (\ref{eq:burgersequation}).
We should regard the weak solutions as the genuine generating function since the integro-differential equation (\ref{eq:inviscidburgers_integral}) is equivalent to (\ref{eq:deformation_coefficientequation}).

A non-differentiable solution to the inviscid Burgers equation can also be addressed by the inviscid limit of the Burgers equation, which is exactly solvable by the Cole-Hopf transformation.
In general, the asymptotic solution of the Burgers equation in the inviscid limit becomes a linear dispersion
$
 f(\lambda,k)
 =
 (k-k_M)/\lambda,
$
where $k_M$ satisfies $f(0,k_M)=0$ and $\partial f(0,k_M)/\partial k>0$.

Specifically, 
the dispersion of the deformed Hamiltonian is the solution of
\begin{align}
 f(\lambda,k)=-2\cos[k-\lambda f(\lambda,k)].
\end{align}
The evolution of the dispersion relation is shown in the inset of Fig.~\ref{fig:deformedcosine} (bottom).
Starting from $f(0,k)=-2\cos k$, the shock wave is formed after $\lambda=1/2$, where the slope at $k=-\pi/2$ diverges, and the dispersion relation converges to $f(\lambda,k)=(k-\pi/2)/\lambda\, (k\in[-\pi/2,3\pi/2])$.

\paragraph{The boost deformation,
  the thermal Drude weight, and the thermal Meissner stiffness}

The thermal Drude weight and the thermal Meissner stiffness 
of the lattice free fermion model
calculated by (\ref{eq:thermaldrudeweight_boostdeformation}) and (\ref{eq:thermalmeissnerstiffness_boostdeformation}) are plotted in Fig.~\ref{fig:deformedcosine}.
As shown in Appendix
\ref{sec:appendix_thermaltransportcoefficients_cleanfermion}, the thermal Drude
weight of a clean fermion system
converges to $\pi vT^2/6$ in the thermodynamic limit $L\to\infty$ at low temperature $T\ll 1$.
The thermal Drude weight (and thermal Meissner stiffness) at $T=0$ is consistent with the CFT result (\ref{eq:cft_thermalmeissnerstiffness}) by taking into account that a complex fermion is equivalent to two real fermions ($c=1/2$) and that $2(c-24h)=-2$ for PBC ($h=1/16$) and $2(c-24h)=1$ for APBC ($h=0$).
However, notice that physical properties of a free fermion depends on the length modulo 4
(for details see Appendix~\ref{sec:freefermion_boosttms}).

\subsection{The XXZ chain with boost deformation}

\begin{figure}[tb]
	\centering
	\includegraphics[width = 0.9\linewidth]{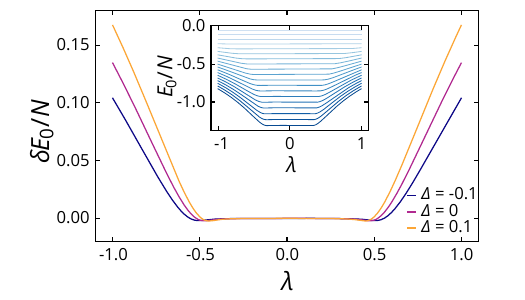}
	\includegraphics[width = 0.9\linewidth]{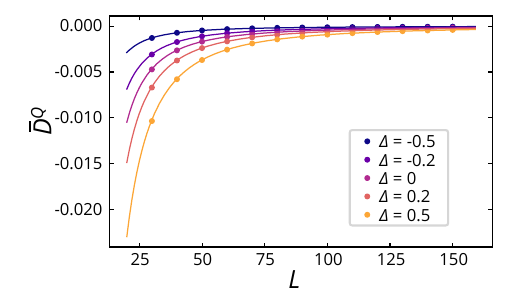}
	\caption{
    (Top)
    The ground state energy
    of the boost-deformed XXZ model with $L = 100$ for $\Delta = -0.1,0,0.1$
    computed from the boost-deformed 
    Bethe ansatz equations.
    Here $\delta E_0= E_0(\lambda) - E_0(0)$.
    The inset shows the ground state energy of the boost deformed XXZ model
    for $ -0.9 \leq \Delta \leq 0.9$ with step size of 0.1 from top to bottom.
    (Bottom)
    The size dependence of the thermal Drude weight
    \eqref{eq:thermaldrudeweight_boostdeformation} at zero temperature
    calculated using the Bethe ansatz equations (dots).
    Solid lines represent
    the CFT predictions \eqref{eq:cft_thermalmeissnerstiffness}. 
		\label{fig:XXZ_ground_state}
	}
\end{figure}

We now turn to the boost deformation of the XXZ model 
\eqref{XXZ Ham}.
As outlined in Sec.\
\ref{Boost boundary condition and deformation in integrable systems},
the boost deformation can be
implemented in
the Bethe ansatz equations. 
Specifically,
we solve 
\begin{equation}
    L[p_1(v_j^{\lambda}) + \lambda h(v_j^{\lambda})] -\sum_{k=1}^{N}p_2(v_j^{\lambda} - v_k^{\lambda}) = 2\pi I_j
    \label{eq:BAE_deform_equ}
\end{equation}
with $p_n(v) = 2 \tan^{-1}\Big(\frac{\tanh{\frac{\gamma v}{2}}}{\tan{\frac{n\gamma}{2}}}\Big)$.
Here, focusing on the ground state at half-filling, $N=L/2$,
the quantum numbers
in \eqref{eq:BAE_deform_equ}
are given by
$I_j = -\frac{N-1}{2}+j-1$. 
We then obtain the
ground state energy $E(\lambda)$
as a function of the boost parameter
(Fig.\ \ref{fig:XXZ_ground_state}).

When $\Delta=0$,
we have checked that the calculation  
using the boost-deformed Bethe ansatz equations
reproduces the free fermion result. 
We observe that for small $\lambda$,
there is a ``plateau-like'' structure,
whereas for larger $\lambda$,  
the ground state energy depends more
sensitively on $\lambda$.
At the free fermion point $\Delta=0$,
this change in the behavior of the ground state
energy coincides with the formation of the
shock wave in the dispersion
at $\lambda = \pm 1/2$.

The finite-size scaling of the zero temperature thermal Drude weight
is shown in the bottom plot of Fig.\ \ref{fig:XXZ_ground_state}.
Here, as we take the limit $\beta\to \infty$
before $L\to \infty$,
the thermal Drude weight and thermal Meissner stiffness
coincide.
We can thus compare the result from
the Bethe ansatz with the CFT prediction.
Recalling \eqref{cft result},
the ground state energy in the presence of boost
at low temperature is
\begin{equation}
    E = E_{\infty} - \frac{c\pi v_s}{6L(1+v^2_s\kappa^2)}.
\label{eq:cft_fenergy}
\end{equation}
Here, 
$c = 1$ is the central charge and $v_s = \frac{\pi \sin\gamma}{\gamma}$ is the sound velocity.
With the identification $i\lambda = \kappa$,
we obtain the CFT prediction
\begin{equation}
        \Bar{D}^{Q} = \frac{1}{2L} \frac{d^2E}{d \lambda^2}\Bigr|_{\lambda = 0} 
        = \Bar{D}^{Q}_{\infty} - \frac{c \pi v_s^3}{6L^2 }.
\label{eq:linear_DW_CFT}
\end{equation}
This is basically the same as \eqref{eq:cft_thermalmeissnerstiffness}.
As shown in Fig.\ \ref{fig:XXZ_ground_state},
the result from the Bethe ansatz
agrees well with the CFT prediction, 
and converges to zero in the $L\to\infty$
limit
as predicted \cite{Kl_mper_2002}.

\begin{figure*}[tb]
  	\includegraphics[width = 0.3\linewidth]{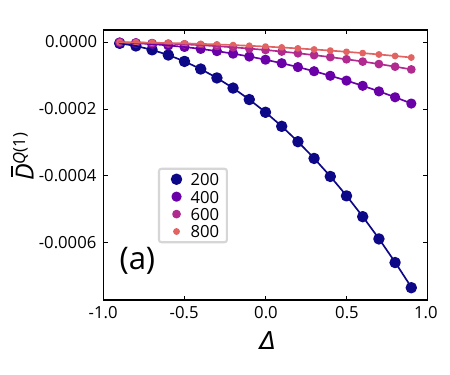}
  	\includegraphics[width = 0.3\linewidth]{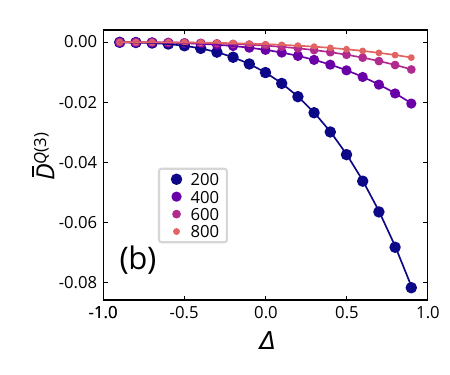}
  	\includegraphics[width = 0.3\linewidth]{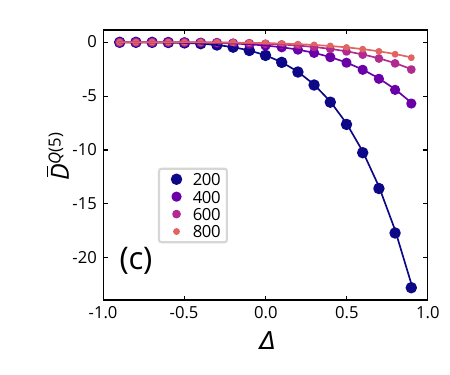}
    \caption{The linear (a),
     third (b), and fifth (c) order nonlinear thermal Drude weights
      at $\lambda = 0$ as defined in Eq.(\ref{eq:nonlinear_dw})
      for different system sizes.
      The dots are from solving the Bethe ansatz and
      the solid lines are calculated from
      Eq.\ \eqref{eq:nonlinear_dw}.
    \label{fig:XXZ_nonlinear_DW} }
\end{figure*}




Using our formalism, it is also possible to
discuss the nonlinear thermal Drude weights.
They can be defined,
following the definition of 
the nonlinear spin Drude weights
\cite{2019arXiv190701212O,2020JSP...tmp..238W},
as
\begin{equation}
	\bar{D}^{Q(n)} = \frac{1}{L}\frac{d^{n+1} E}{d\lambda^{n+1}}\Bigr|_{\lambda = 0},
	\quad 
	n>1.
\label{eq:nonlinear_dw}
\end{equation}
The results are shown in Fig.\ \ref{fig:XXZ_nonlinear_DW}.
From Fig.\ \ref{fig:XXZ_nonlinear_DW},
we see that
the CFT prediction still fits well 
the higher order nonlinear thermal
Drude weight obtained from the Bethe ansatz,
if we assume $\Bar{D}^{Q(n=3,5)}_{\infty} = 0$:
the nonlinear thermal Drude weights
also converge to zero at large system sizes.

\begin{figure*}[tb]
  	\includegraphics[width = 0.3\linewidth]{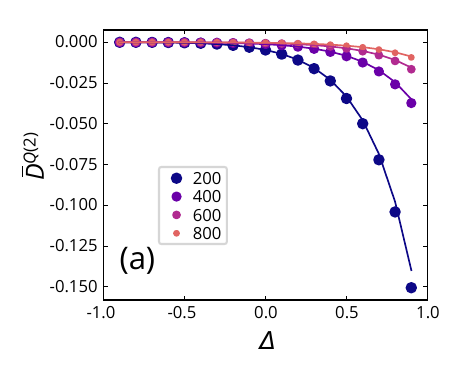}
  	\includegraphics[width = 0.3\linewidth]{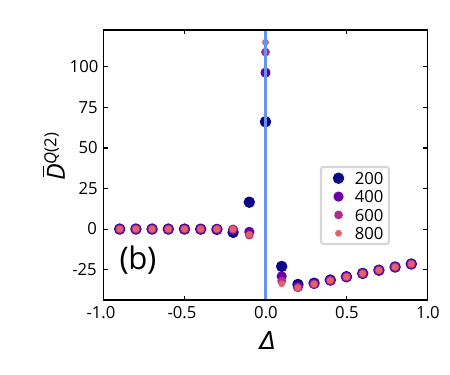}
  	\includegraphics[width = 0.3\linewidth]{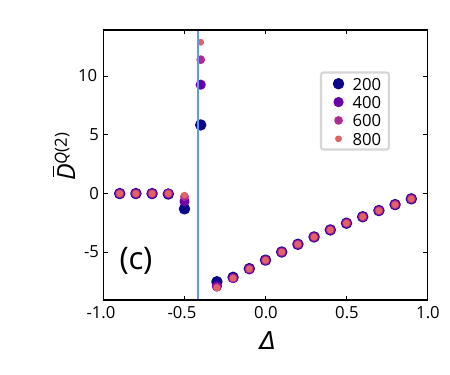}
    \caption{The second-order nonlinear thermal Drude weight
      at $\lambda = 0.2$ (a),
      $\lambda = 0.5$ (b), and $\lambda = 0.7$ (c)
      for different system sizes. The dots are from
      solving the Bethe ansatz and
      the solid lines are calculated from Eq.\ \eqref{eq:dq_finite_lam}. The blue vertical lines in (b) and (c) indicate the divergence region calculated from Eq.\ \eqref{eq:dq_finite_lam}. 
    \label{fig:XXZ_nonlinear_DW_finite_lam} }
\end{figure*}

Finally,
we can also obtain the nonlinear thermal Drude weights
at finite boost parameter $\lambda$,
as shown in Fig.\ \ref{fig:XXZ_nonlinear_DW_finite_lam}.
The nonlinear thermal Drude weights at zero temperature could be computed as
\begin{equation}
\Bar{D}^{Q{(n)}} (\lambda) 
=  \frac{1}{L}\frac{d^{n+1} E}{d \lambda^{n+1}}  \propto \frac{1}{1-v^2_s\lambda^2}.
\label{eq:dq_finite_lam}
\end{equation}
Since $v_s = \frac{\pi \sin\gamma}{\gamma} \in (0,\pi)$,
as we change $\Delta$,
there is a singularity
at $v_s=1/\lambda$.
We indeed see
in our Bethe ansatz calculation 
that at certain value of $\Delta$,
$\Bar{D}^{Q{(n)}} (\lambda)$ diverges
for $\lambda = 0.5$ and $\lambda = 0.7$
(Fig.\ \ref{fig:XXZ_nonlinear_DW_finite_lam}).
We confirmed that these divergence values
coincide with $v_s = 1/\lambda$.

These findings should be compared 
with the behaviors of
the nonlinear spin Drude weights
\cite{2019arXiv190701212O,
PhysRevB.103.L201120, tanikawa2021fine}.
First, we did not observe divergences for
$\bar{D}^{Q(n=3,5)}$
in contrast with the nonlinear spin Drude weights.
Second, 
the Bethe ansatz results 
for $\bar{D}^{Q(n=3,5)}$
are described very well by  
the CFT predictions.  

To address these questions (at least partially), 
let us focus on the non-interacting case 
and consider the effect of
the non-linearity of the dispersion
on the nonlinear thermal Drude weight. 
\footnote{
We thank Hosho Katsura who suggested
this calculation.
}
We consider the single particle spectrum:
\begin{equation}
     \epsilon(p) = v_1 p + v_3 p^3 + \cdots = \sum_{m=1} v_{2m-1} p^{2m-1}.
\end{equation}
As in Eq.\ \eqref{twisted k quantization},
we impose the energy-twisted boundary 
condition,
\begin{equation}
     p - \lambda \sum_{m=1} v_{2m-1} p^{2m-1} = \frac{2\pi}{L} 
     (-r+\alpha),
     \quad 
     r \in \mathbb{Z}
\end{equation}
where $\alpha=0(1/2)$ for PBC (APBC).
This quantization condition on $p$
can be solved order-by-order 
in $\lambda$.
If we expand the momentum $p$ as
$
    p = \sum_{l=0} \lambda^l A_l,
$
we can determine $A_l$ as 
\begin{align}
    \begin{split}
        A_0 = & \frac{2\pi}{L} 
        (-r+\alpha),
        \\
        A_1 = &  \sum_{m=1} v_{2m-1} A_0^{2m-1},
        \\
        A_2 = & \sum_{m=1}  v_{2m-1} (2m-1) A_0^{2m-2}A_1,
        \\
        \vdots\\
        A_l = &  \sum_{m=1} v_{2m-1} \sum_{\substack{i_1<i_2< \cdots\leq l-1 \\
        r_1 + r_2 + \cdots = 2m-1 \\
        i_1 r_1 + i_2 r_2 + \cdots = l-1}}  C_{r_1, r_2\cdots}A_{i_1}^{r_1}A_{i_2}^{r_2}A_{i_3}^{r_3} \cdots,  
        \\
        \vdots
    \end{split}
\end{align}
with $C_{r_1, r_2\dots} =\frac{(2m -1)!}{r_1! r_2! \dots}$. 
We assume the ground state 
where all 
single-particle states with $-r+\alpha<0$ are filled. 
The ground state energy 
is then given by
$
E(\lambda) =  \sum_{
r= 1 }^{\infty}
\sum_{m=1} v_{2m-1} p^{2m-1}(\lambda)
=  
\sum_{r=1}^{\infty}  \sum_{m=1} v_{2m-1}  
\left( \sum_{l=0} \lambda^l A_l\right)^{2m-1}. 
$
The nonlinear thermal
Drude weight is obtained 
by taking the (higher) 
derivative 
of the ground state energy 
with respect to $\lambda$.
Focusing on
the contributions
from the linear part
of the dispersion,
\begin{align}
\frac{d^{n} E}{d \lambda^{n}}
=
\frac{n! v_1^{n+1} 2\pi}{L}
\sum_{r=1}^{\infty} 
(-r+\alpha)
+\cdots
\end{align}
We have so far 
focused on 
the left-movers.
Combining 
the contributions 
from the right-movers, 
for which 
the dispersion is given by
$
\sum_m v_{2m-1}(-p)^{2m-1}
$,
we see that 
the contributions 
cancel for odd $n$,
while they add up for 
even $n$.
For APBC, we can regularize
$\sum_{r=1}^{\infty} (-r +\alpha)
= -1/24
$.
Hence, 
\begin{align}
\frac{d^{n} E}{d \lambda^{n}}
&=
\left\{
\begin{array}{ll}
\displaystyle
-\frac{n! v_1^{n+1} \pi}{6L}
& n: \mbox{even}
\\
0 & n: \mbox{odd}
\end{array}
\right.
\end{align}
This is consistent with the calculation from Eq.\ \eqref{cft result} 
which suggests
\begin{equation}
    E(\lambda) = 
     -\frac{ (c- 24 h) \pi v }{ 6L(1 - v^2 \lambda^2)} 
\end{equation}
(we take $c = 1$ and $h = 0$). 
The nonlinear thermal
Drude weight is then
\begin{align}
\bar{D}^{Q(n)} = & \frac{1}{L}\frac{d^{(n+1)} E}{d \lambda^{(n+1)}} 
=
\left\{
\begin{array}{ll}
\displaystyle
-\frac{(n+1)! v_1^{n+2} \pi}{6L^2}
& n: \mbox{odd}
\\
0 & n: \mbox{even}
\end{array}
\right.
\end{align}

To conclude,
we see that the leading contributions
to the nonlinear thermal Drude weights
come from the linear part of the dispersion $v_1$.
This should be contrasted with  the nonlinear spin Drude weights,
which are governed by the non-linearity of
the dispersion, $v_{2m-1>1}$
\cite{katsura}.
I.e., the purely linearly-dispersing band
or CFT predicts   
vanishing nonlinear spin Drude weights
and fails to reproduce lattice calculations.
On the other hand, for the nonlinear thermal
Drude weight, CFT still captures the dominant 
contributions.

\section{Conclusion}
\label{Conclusion}

We have formulated
a symmetry twist of the boundary condition relevant to thermal transport
as the energy-twisted boundary condition, and shown that the stiffness against the twist quantifies thermal transport properties. 
We have also identified its bulk
counterpart as the boost deformation, which has been studied in the context of a long-range deformation of integrable systems.
These have a close analogy with the $U(1)$ twisted boundary condition and the equivalent bulk $U(1)$ gauge transformation relevant to electric transport.
The relations have been confirmed by the agreement of the thermal Drude weight and the thermal Meissner stiffness estimated by each method. Specifically, the CFT result under the energy-twisted boundary condition agrees with the other results as far as CFT is applicable. 
A rigorous relation between the stiffnesses and the ac conductivity is shown only at the free fermion point.

The energy-twisted boundary condition is imposed on tori and is mostly suited for the evaluation of the partition function via the reconnection of tensor networks.
It is thus compatible in particular with exact methods in $1+1$ dimensions and numerical analysis in any dimensions.
We have demonstrated how this method works in the estimation of the thermal Meissner stiffness of CFTs based on the modular transformation, and also that of the transverse-field Ising model and disordered lattice fermions in $1+1$ dimensions based on the transfer matrix.

The boost deformation is a sort of integrable deformation applied in the bulk, and thus suited for integrable systems in $1+1$ dimensions. 
We showed an implementation of the boost deformation in the Bethe ansatz,
and addressed the linear and nonlinear thermal Drude weights of the XXZ
Heisenberg spin chain.
We also analyzed the energy-twisted deformation of the free fermion chain
via the inviscid Burgers equation. The agreement of the thermal Meissner stiffness with that of the Ising CFT under the energy-twisted boundary condition indicates an equivalence of the bulk and boundary-condition methods at least in a critical model.

Extending these analyses to a wider range of systems,  
beyond those studied in this paper, is an important open question.
In particular, unlike the energy-twisted boundary condition,  
the boost deformation is formulated
by making use of the integrability of (1+1)D quantum many-body systems,
or in continuum systems with Lorentz invariance.
It is important to formulate and study the boost deformation
outside of these contexts. 
Also interesting is to
study the energy-twisted boundary condition and boost deformation in
quantum many-body systems in higher dimensions.
As a simple warm-up,
in Appendix \ref{Quantum Hall system with boost deformation},
we present the implementation of the energy-twisted boundary condition
in the 2d integer quantum Hall effect.
Just like Laughlin's argument for the quantized Hall conductance,
the transverse energy transport can be induced by
an adiabatic change in the boost parameter.
Studying interacting 2d quantum many-body systems
(e.g., fractional quantum Hall systems)
using the energy-twisted boundary condition and boost deformation
would be a natural next step.
In this regard, 
it would be interesting to 
make a comparison with
other formalisms, 
such as 
Ref.\ \cite{2020PhRvB.101d5137K}.


  \appendix

\section*{Acknowledgments}

We thank Vir Bulchandani, 
Hosho Katsura, Jonah Kudler-Flam, Kentaro Nomura, and Kiyohide Nomura for discussions.
R.N.~is supported by JSPS KAKENHI Grant No. JP17K17604 and JST CREST Grant No. JPMJCR18T2.
S.R.~is supported by the National Science Foundation under 
Award No.\ DMR-2001181, and by a Simons Investigator Grant from
the Simons Foundation (Award No.~566116).
This work is supported by the Gordon and Betty Moore Foundation 
through Grant GBMF8685 toward the Princeton theory program.
\textcolor{black}{This work was performed in part at Aspen Center for Physics, which is supported by National Science Foundation grant PHY-1607611. 
This work was partially supported by a grant from the Simons Foundation.}

\section{Graviphoton field on 2-torus}
\label{Graviphoton field on 2-torus}

Consider the (1+1)D Euclidean spacetime with the metric
\begin{align}
 ds^2
 =
 (d\tau+A^\text{E}_x dx)^2
 +
 dx^2,
 \label{eq:euclidean_spacetime}
\end{align} 
where $A^{\mathrm{E}}_x$ is the background gravitomagnetic
vector potential.
By the Wick rotation,
$\tau=it$ and $A_x^\text{E}=iA_x^\text{g}$,
the line element in the Minkowski signature is given by 
$
 ds^2
 =
 -(dt+A^\text{g}_x dx)^2
 +
 dx^2.
$
The gravitomagnetic vector potential
induces a gravitational counterpart of
magnetic flux.
Provided that the gravitomagnetic vector potential $A^\text{E}_x$ is static,
the metric \eqref{eq:euclidean_spacetime}
is obtained from the regular flat metric
by a transformation
\begin{align}
 (\tau,x)
 \to
 (\tau+\beta a(x), x),
 \label{eq:spacetime_transformation}
\end{align}
where $a(x)=\beta^{-1}\int_0^x dx'A_x^\text{E}(x')$.
To be consistent with the spatial periodicity,
we assume $A_x^{\mathrm{E}}$ is a periodic function of $x$,
$A_x^{\mathrm{E}}(x+L)=
A_x^{\mathrm{E}}(x)$. 
If we start from 
the spacetime 2-torus with periodicity
\begin{align}
 (\tau,x)
 \sim
 (\tau+\beta,x)
 \sim
 (\tau,x+L),
 \label{eq:periodicity_before}
\end{align}
then after the transformation
the new identification condition is given by
\cite{2015arXiv151202607G}
\begin{align}
 (\tau,x)
 \sim
 (\tau+\beta,x)
 \sim
 (\tau+\beta a(L),x+L).
 \label{eq:periodicity_after}
\end{align}

\section{Lattice spin systems and transfer matrix formalism}
\label{Lattice spin systems and transfer matrix formalism}

In this appendix,
we review the derivation of the column-to-column transfer matrix of the
transverse-field Ising model following
\cite{10.1143/PTP.56.1454,
PhysRevB.31.2957,
10.1143/PTP.78.787,
pirvu10,rams15},
and derive the twisted partition function.
Consider the transverse-field Ising model in a general form
\begin{align}
 H
 =
 \sum_{i=1}^L
 \left(
 -J_{i}\sigma_i^x\sigma_{i+1}^x
 -
 h_i \sigma_i^z
 \right),
\end{align}
satisfying PBC ($\sigma_{L+1}=\sigma_1$).
By Trotterizing the imaginary time direction,
the partition function is written
in terms of the row-to-row transfer matrix $V$ as
$
 Z
 =
 \text{Tr}\, e^{-\beta H}
 \simeq
 \text{Tr}\,V^{M}
$,
where an integer $M$ is the length of the temporal direction.
The transfer matrix
can be written
as a product form,
$V={V^{(1)}}^{1/2}V^{(2)}{V^{(1)}}^{1/2}$,
where
\begin{align}
 &V^{(1)}
 =
 \prod_{i=1}^L
   e^{
 \gamma_{i}\sigma_i^z
 },
   \quad
 V^{(2)}
 = 
 \prod_{i=1}^L
   e^{
 K_{i}\sigma_i^x\sigma_{i+1}^x
   }.
\end{align}
Here,
the coefficients are defined by $K_{i}=\beta J_{i}/M$ and $\gamma_i=\beta h_i/M$.
By introducing vectors $B_0(\epsilon)=(\sqrt{\cosh\epsilon},0)^T$ and $B_1(\epsilon)=(0,\sqrt{\sinh\epsilon})^T$, we obtain \cite{pirvu10}
\begin{align}
 &e^{K_{i}\sigma_i^x\sigma_{i+1}^x} =\notag\\
 &\sum_{s_i,t_{i+1}}
 B_{s_i}^T(K_{i})
 B_{t_{i+1}}(K_{i})
 \left(\sigma_i^x\right)^{s_i}
 \left(\sigma_{i+1}^x\right)^{t_{i+1}},
 \label{eq:appendix_exponentialisingcoupling}
\end{align}
where $s_i$ and $t_{i+1}$ take $0,1$,
and thus
\begin{align}
 V^{(2)}
 &=
 \sum_{s_1,t_1,k_2,\cdots,k_L}
 B_{s_1}^T(K_{1})
 C^{k_2}_2\cdots C^{k_L}_L B_{t_1}(K_{L}) \notag\\
  &\quad
    \times
 \left(\sigma_1^x\right)^{s_1+t_1}
 \otimes
 \left(\sigma_2^x\right)^{k_2}
 \cdots
 \otimes
 \left(\sigma_L^x\right)^{k_L},
\end{align}
where
$
 C_i^k
 =
 \sum_{s}
 B_{s}(K_{i-1})
 B_{s+k}^T(K_{i})
$
and, the subscript of $B_s(\epsilon)$ is defined modulo 2.
By making the imaginary-time coordinate explicit, we obtain $V^M=\prod_{j=1}^M V_j$ where
\begin{align}
 V_j&=
 \sum_{s_{1j},t_{1j},k_{2j},\cdots,k_{Lj}}
 B_{s_{1j}}^T(K_{1})
      C^{k_{2j}}_{2j}\cdots C^{k_{Lj}}_{Lj} B_{t_{1j}}(K_{L})
      \notag\\
 &\quad \times
 X_{1j}^{s_{1j}+t_{1j}}
 \otimes
 X_{2j}^{k_{2j}}
 \cdots
 \otimes
 X_{Lj}^{k_{Lj}},
 \label{eq:transverseising_transfermatrix}
\end{align}
and 
$
 X_{ij}^{k}
 =
 e^{\gamma_{i}\sigma_{i}^z/2}
 (\sigma_{i}^x)^k
 e^{\gamma_{i}\sigma_{i}^z/2}
$.

When the spacetime is twisted by $a$ lattice sites, Ising coupling connects the boundary spin at a position $(L,j)$ to the spin on the other side at $(1,j+a)$.
This changes the ket vector in (\ref{eq:transverseising_transfermatrix}) as $B_{t_{1j}}(K_{L})\to B_{t_{1j+a}}(K_{L})$.
Inserting the identity matrix $\mathbb{1}=\sum_{\tau_j}|\tau_j\rangle\langle\tau_j|$ of the auxiliary 2-dimensional space in front of $B_{t_{1j+a}}(K_{L})$, the transfer matrix on a twisted spacetime becomes
\begin{align}
 V_j(a)&=
 \sum_{\tau_j}\sum_{k_{1j},\cdots,k_{Lj}}
 \langle\tau_{j-a}|
 C^{k_{1j}}_{1j}\cdots C^{k_{Lj}}_{Lj}
 |\tau_j\rangle \notag\\
       &\quad
         \times
 X_{1j}^{k_{1j}}
 \otimes
 X_{2j}^{k_{2j}}
 \cdots
 \otimes
 X_{Lj}^{k_{Lj}}.
\end{align}
Due to the duality between $C$ and $X$, the column-to-column transfer matrix is
\begin{align}
 W_i
 &=
 \sum_{\sigma_i}\sum_{k_{i1},\cdots,k_{iM}}
 \langle\sigma_{i}|
 X_{i1}^{k_{i1}}\cdots X_{i1}^{k_{iM}}
   |\sigma_i\rangle
   \notag \\
 &\quad \times
 C_{i1}^{k_{i1}}
 \otimes
 C_{i2}^{k_{i2}}
 \otimes
 \cdots
 \otimes
 C_{iM}^{k_{iM}},
\end{align}
which satisfies $Z=\sum_{\sigma}\langle\sigma_1\cdots\sigma_L|\prod_jV_j(a)|\sigma_1\cdots\sigma_L\rangle=\sum_{\tau}\langle\tau_{1-a}\cdots\tau_{M-a}|\prod_iW_i |\tau_1\cdots\tau_M\rangle$.
Here, the auxiliary spin is also periodically identified: $\tau_{j+M}=\tau_j$.

The spin operators $C$ and $X$ can be rewritten by a similar expression as the original $X$ and $C$, respectively, as
\begin{align}
 C_{ij}^k&=
 \alpha_i
 e^{K^\ast_{i-1}\tau^z_{j}/2}
 \left(\tau_{j}^x\right)^k
 e^{K^\ast_{i}\tau^z_{j}/2}, \\
 X_{ij}^k&=
 \beta_i\sum_{s}B_s({\gamma_i^\ast})B_{s+k}^T({\gamma_i^\ast}),
 \label{eq:transverseising_x}
\end{align}
where $\tanh K_{i}=e^{-2K_{i}^{\ast}}$ and $\tanh \gamma_i=e^{-2\gamma_i^\ast}$ from the standard notation \cite{schultz64}, $\alpha_i=(\sinh 2K_{i-1}\sinh 2K_{i}/4)^{1/4}$, and $\beta_i=(2\sinh 2\gamma_i)^{1/2}$. 
Notice that the vector $B_s(\epsilon)$ in (\ref{eq:transverseising_x}) is the spinor of the real spin $\sigma$, while that in (\ref{eq:appendix_exponentialisingcoupling}) is of the auxiliary spin $\tau$.
\textcolor{black}{These expressions lead to} the column-to-column transfer matrix in terms of the auxiliary spin as $W_i=(\alpha_i\beta_i)^M{W_{i-1}^{(1)}}^{1/2}W_i^{(2)}{W_{i}^{(1)}}^{1/2}$, where
\begin{align}
 &W_{i}^{(1)}
 =
   \prod_{j=1}^M
   e^{K_{i}^{\ast}\tau_j^z},
   \quad
 W_i^{(2)}
 =
   \prod_{j=1}^M
   e^{\gamma_i^{\ast}\tau_j^x\tau_{j+1}^x}.
\end{align}

The transfer matrix is diagonalized by introducing fermionic representation via the Jordan-Wigner transformation:
$
 \tau_j^z
 =
 2c_j^{\dagger}c_j-1,
 $
 $
 \tau_j^+
 =
 \tau_j^x+i\tau_j^y
 =
 2e^{i\pi\sum_{l<j}c_l^{\dagger}c_l}c_j^{\dagger}.$
The Ising coupling is then written by the hopping of the Jordan-Wigner fermions as
\begin{align}
 &\tau_j^x\tau_{j+1}^x
 =
 (c_j^{\dagger}-c_j)(c_{j+1}^{\dagger}+c_{j+1}),
\end{align}
where, at the boundary, $c_{M+1}=-e^{i\pi\sum_jc_j^{\dagger}c_j}c_1$ is imposed.
Since the Hamiltonian is bilinear in the fermion operators, the total fermion number $F=\sum_{j}c_j^{\dagger}c_j$ modulo 2 is conserved.
The Fock space is then decomposed into even- and odd-fermion-number subspaces, within which the fermion operator obeys APBC and PBC, respectively.
The boundary condition in the temporal direction appears in the frequencies of the Fourier mode:
\begin{align}
 c_j
 =
 \frac{e^{-i\pi/4}}{\sqrt{M}}
 \sum_{\omega}
 e^{i\omega j}
 c_\omega,
\end{align}
where
\begin{align}
 \omega
 =
 \left\{
 \begin{array}{lr}
  \displaystyle
  \pm\frac{\pi}{M}, \pm\frac{3\pi}{M}, \cdots, \pm\frac{(M-1)\pi}{M}\,
  &(\text{$F$:even})\\[+7pt]
  \displaystyle
  0, \pm\frac{2\pi}{M},  \cdots, \pm\frac{(M-2)\pi}{M}, \pi\,
  &(\text{$F$:odd})
 \end{array}
 \right.
 \label{eq:xy_fouriermodes1}
\end{align}
for even $M$, and
\begin{align}
 \omega
 =
 \left\{
 \begin{array}{lr}
  \displaystyle
  \pm\frac{\pi}{M}, \pm\frac{3\pi}{M}, \cdots, \pm\frac{(M-2)\pi}{M},\pi\,
  &(\text{$F$:even})\\[+7pt]
  \displaystyle
  0, \pm\frac{2\pi}{M}, \cdots, \pm\frac{(M-1)\pi}{M}\,
  &(\text{$F$:odd})
 \end{array}
 \right.
 \label{eq:xy_fouriermodes2}
\end{align}
for odd $M$.
The transfer matrix is then written as
\begin{align}
 W_i
 =
 \prod_{\omega\in[0,\pi]}W_i(\omega),
 \label{eq:ising_transfermatrix_fourier}
\end{align}
where the summation is over the non-negative part of (\ref{eq:xy_fouriermodes1}) and (\ref{eq:xy_fouriermodes2}), and by using $n_\omega=c_\omega^{\dagger}c_\omega^{\ }$,
\begin{align}
 W_i(0)
 &=
   e^{
 (K^{\ast}_{i-1}+2\gamma_i^{\ast}+K^{\ast}_{i})(n_0-1/2)},
 \label{eq:xytransfermatrix0_jordanwigner}\\
 W_i(\pi)
 &=
   e^{
 (K^{\ast}_{i-1}-2\gamma_i^{\ast}+K^{\ast}_{i})(n_\pi-1/2)},
 \label{eq:xytransfermatrixpi_jordanwigner}\\
 W_{i}(\omega)
 &=
   e^{
   K^{\ast}_{i-1}(n_\omega+n_{-\omega}-1)}
    \notag\\
  &\quad \times
    e^{
 2\gamma_i^\ast
 \left(\cos\omega(n_\omega+n_{-\omega})
 +\sin\omega(c_{-\omega}^{\dagger}c_\omega^{\dagger}+c_{\omega}^{\ }c_{-\omega}^{\ })\right)
    }
    \notag\\
  &\quad \times
    e^{
    K^{\ast}_{i}(n_\omega+n_{-\omega}-1)},
 \label{eq:xytransfermatrix2_jordanwigner}
\end{align}
We can decompose the Fock space into subspaces specified by Fourier components of $\omega=0$, $\pi$, and combined $\omega$ and $-\omega$.
To be specific, the $\omega=0/\pi$ subspace is spanned by $|0\rangle$ and $c_{0/\pi}^\dagger|0\rangle$, and a $\omega\neq 0,\pi$ subspace by $|0\rangle$, $c_{\omega}^\dagger|0\rangle$, $c_{-\omega}^\dagger|0\rangle$, and $c_{-\omega}^\dagger c_{\omega}^\dagger|0\rangle$.
The trace of the column-to-column transfer matrix (\ref{eq:ising_transfermatrix_fourier}) is thus \textcolor{black}{the product of the traces of} small matrices corresponding to the subspaces.

When the spacetime is twisted, the fermion operators at $i=1$ are changed as $c_{m}\to c_{m+a}$, which shifts the Fourier mode by a frequency-dependent phase as
\begin{align}
  c_{\omega}
 \to
 e^{ia\omega}
  c_{\omega}.
\end{align}
This modifies the trace operation so that the bra vector is shifted by a phase determined by the number of fermion and the frequency as
\begin{align}
 &\langle n_{0}|
 \to
 \langle n_{0}|,\quad
 \langle n_{\pi}|
 \to
 (-1)^{an_\pi}
 \langle n_{\pi}|, \notag\\
 &\langle n_{\omega}n_{-\omega}| \to
 e^{ia\omega(n_{\omega}-n_{-\omega})}
 \langle n_{\omega}n_{-\omega}|.
 \label{eq:xy_twistbasis}
\end{align}
Finally, the partition function after the twist is the sum of contributions from even- and odd-fermion-number spaces as
\begin{align}
 Z(a)
 \propto
 \sum_{\pm}
 \prod_{\omega\in[0,\pi]}
 \text{Tr}'_\omega
 \left[
 \frac{1\pm(-1)^F}{2}
 \prod_{i=1}^{L} 
 W_{i}(\omega)
 \right],
\end{align}
where the first summation is over the even- and odd-fermion-number spaces, and $\text{Tr}'_\omega$ is \textcolor{black}{the trace over a $\omega$ subspace} with the modified bra vector (\ref{eq:xy_twistbasis}).
A proportionality constant $(\alpha_i\beta_i)^{LM}$ is omitted.
Specifically, the trace of a twisted $\omega\neq 0$ subspace  is
\begin{align}
 &\text{Tr}'_\omega
 \left[
 (\pm 1)^F
 \prod_{i=1}^{L} 
 W_{i}(\omega)
 \right] \notag\\
 & =
 \begin{pmatrix}
  \langle 0| \\
  \pm e^{ia\omega}\langle 0| c_{\omega} \\
  \pm e^{-ia\omega}\langle 0| c_{-\omega} \\ 
  \langle 0| c_{\omega} c_{-\omega}
 \end{pmatrix}^T
 \prod_{i=1}^{L} 
 W_{i}(\omega)
 \begin{pmatrix}
  |0\rangle \\
  c_{\omega}^\dagger|0\rangle \\
  c_{-\omega}^\dagger|0\rangle \\
  c_{-\omega}^\dagger c_{\omega}^\dagger|0\rangle
 \end{pmatrix}.
\end{align}
The matrix element of the Fourier-decomposed transfer matrix $W_i(\omega)$ can be found in \cite{schultz64}.

\section{Transfer matrix method for free fermion models}
\label{Transfer matrix method for free fermion models}


Following \cite{Yang_2009}, we consider the free fermion model
on a 1d lattice with the Hamiltonian
\begin{equation}
	H = \sum_i h_{i,i+1} = -\sum_{i} t_i (c_i^{\dagger}c_{i+1}+h.c.) +\sum_i (U_i - \mu)c_i^{\dagger}c_i.
\end{equation}
To implement the transfer matrix method, 
we decompose the system into even and odd sites and define 
$H_1 = \sum_{i = \text{odd}} h_{i,i+1}$,
$H_2 = \sum_{i = \text{even}} h_{i,i+1}$.
With the local transfer matrices defined as
$
  V_1 = e^{-\epsilon H_1} = \prod_{i = \text{odd}} v_{i,i+1},
$
and
$
V_2 = e^{-\epsilon H_2} = \prod_{i = \text{even}} v_{i,i+1},
$
where $v_{i,i+1} = e^{-\epsilon h_{i,i+1}}$, 
the partition function can be written as 
\begin{equation}
	Z = \text{Tr}\, (e^{-\beta H})
  = \text{Tr}\, (V_1V_2)^{M}+O(\epsilon^2),
\end{equation}
where $\epsilon = \beta/M$ and $M$ is Trotter number. 
By inserting the complete set of states,
we can write the row-to-row partition function as 
\begin{equation}
	Z =
  \sum_{\{n_{i}^{l}\}}\prod_{l = 1}^{M}(v_{1,2}^{2l-1,2l}\cdots v_{N-1,N}^{2l-1,2l})(v_{2,3}^{2l,2l+1}\cdots v_{N,1}^{2l,2l+1}),
\end{equation}
where $v_{i,i+1}^{l,l+1} = \langle n_i^l,n_{i+1}^l| v_{i,i+1}|
n_i^{l+1},n_{i+1}^{l+1}\rangle$ with $i$ and $l$ represent the site number in
space and Trotter directions, respectively. 

In order to go from the row-to-row
to column-to-column transfer matrix, we rotate each block as
\begin{equation}
	\tau_{i,i+1}^{l,l+1} = \langle n_i^l,1-n_{i}^{l+1}| v_{i,i+1}| 1-n_{i+1}^{l},n_{i+1}^{l+1}\rangle.
\end{equation}
Explicitly, it is given by 
\begin{equation}
	\tau_{i,i+1}^{l,l+1} = b_i 
	\begin{pmatrix}
		u_i& 0&0&0\\
		0 & a_i-w_i&b_i^{-1}&0\\
		0&b_i&a_i+w_i&0\\
		0&0&0& u_i
	\end{pmatrix}
\end{equation}
in the basis of $\{|00\rangle, |01\rangle, |10\rangle, |11\rangle\}$
with parameter defined as 
\begin{equation}
	\begin{split}
		&\alpha_i = \frac{-\epsilon(U_i-\mu)}{2},
    \quad \gamma_i = \sqrt{\alpha_i^2+\epsilon^2t_i^2},
    \\
		&b_i = e^{\alpha_i}, \quad u_i = \frac{\epsilon t_i \sinh\gamma_i}{\gamma_i},
    \\
		&a_i = \cosh\gamma_i, \quad w_i = \frac{\alpha_i \sinh\gamma_i}{\gamma_i}.
	\end{split}
\end{equation}
Therefore, the partition function
in terms of the column-to-column transfer matrices
is written as 
\begin{equation}
	Z = \text{Tr}\, [T_{1,2}T_{2,3}\cdots T_{N,1}],
  \quad
	T_{i,i+1} = \prod_{l}\tau_{i,i+1}^{l,l+1}.
\end{equation}

Once we write the partition in the matrix form,
we can perform the Fourier transform in the Trotter direction,
and the partition function
can be written as
\begin{align}
  &
  Z = \prod_{i}C_{i}^M \prod_{\omega}\text{Tr}\, [2+T_{\omega}],
    \,\,\,
    T_{\omega} = \prod_{i = 1}^{N/2} t_{2i-1}t_{2i,\omega},
\end{align}
where $C_i = u_ib_i$,
and $t_{2i-1}$ and $t_{2i,\omega}$ are defined as 
\begin{align}
  t_{2i-1} &= \frac{1}{u_{2i-1}}
             \begin{pmatrix}
               a_{2i-1} - w_{2i-1}& b_{2i-1}^{-1} \\
               b_{2i-1} & a_{2i-1} + w_{2i-1}
             \end{pmatrix},
  \\
  t_{2i} &= \frac{1}{u_{2i}}\begin{pmatrix}
    a_{2i} + w_{2i} & e^{-i\omega}b_{2i} \\
    e^{i\omega}b_{2i}^{-1} & a_{2i} -w_{2i}
	\end{pmatrix}.
\end{align}
We use the even number of Trotter sites,
and hence $\omega = \frac{(2m+1)\pi}{M}$ with $m = -\frac{M}{2}, \cdots -1,0,\cdots \frac{M}{2}-1$.

\subsection{Phase-twisted boundary condition}
\begin{figure*}[t]
	\centering
	\includegraphics[width = 0.32\linewidth]{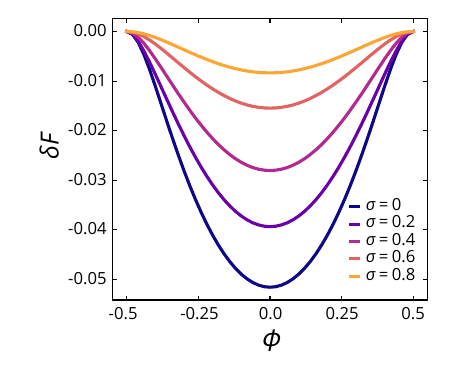}
  \includegraphics[width=0.32\linewidth]{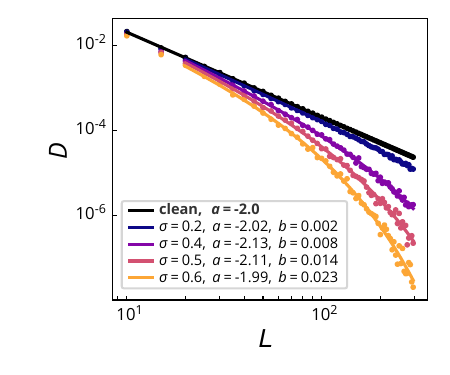}
  \includegraphics[width=0.32\linewidth]{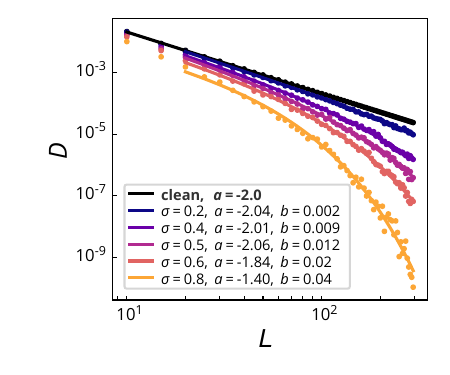}
	\caption{
    (Left)
    The $U(1)$-phase-twist variation of the free energy of
    the free fermion model with random on-site potential.
    The curves are shown as varying $\sigma$ from $0$ to $0.8$. The calculation is done with $L = 50$ and $\beta = 100$. 
    The electrical Meissner stiffness for
    the disordered free fermion chain
    with random on-site potential (middle) 
    and 
    with random hopping (right).
    Dots represents the results from the transfer matrix method
    and solid lines are fitting with $D \sim L^a e^{-bL}$. The fitting
    parameter are labeled in plots with $\beta = 500$.
		\label{fig:fermiontwist}
	}
\end{figure*}

Now we consider the system with phase twisted boundary condition, i.e.,
$t_N \rightarrow t_N e^{2\pi i \phi}$ where $\phi = \Phi/\Phi_0$.
This results in the change of the parameter $u_N$
in the transfer matrix $\tau_{N,1}$,
\begin{equation}
	\tau_{N,1}^{l,l+1} = b_N 
		\begin{pmatrix}
		u_N& 0&0&0\\
		0 & a_N-w_N&b_N^{-1}&0\\
		0&b_N&a_N+w_i&0\\
		0&0&0& u_N^{*}
	\end{pmatrix}.
\end{equation}
Accordingly,
the modified partition function is 
\begin{equation}
	Z = \prod_{i} C_i^M\prod_{\omega} \text{Tr}\, [2\cos(2\pi \phi)+T_{\omega}]. 
\end{equation}
We use $M = 2N$ to ensure the convergence of the partition function. The results
are shown in Fig.\ \ref{fig:fermiontwist}.
The period of $\phi$ is 1 which is equal to a phase twist of $2\pi$. Here we consider the system with onsite random potential $U_i$ to be Gaussian distributed with variance $\sigma$. \textcolor{black}{We see that as the disorder strength increases, the free energy curves become more flat} which means the system is more localized and less sensitive to boundary conditions.

Then we compute the electrical Meissner stiffness 
$
D = (2L)^{-1}d^2 F/d\phi^2
$
(Fig.\ \ref{fig:fermiontwist}).
For the clean free fermion system (black),
we observe
the electrical Meissner stiffness decays algebraically as $D \sim L^{-2}$
for $L\ll \beta$,
which we confirmed
is consistent with the analytical result.
(Here, we take the parameter $\beta = 500$.)
On the other hand,
for the high temperature (long wire) regime, $\beta \ll L$,
the Meissner stiffness decays exponentially.

We also studied
two types of disordered fermion chains, 
one with on-site disorder,
and the other with bond disorder.
Here, for the on-site randomness, 
we consider $U_i$ to be Gaussian distributed with
variance $\sigma$.
For the random hopping model,
the hopping amplitudes are drawn from a uniform distribution,
$t_{i,i+1} \in [1-\sigma,1+\sigma]$. 
We focus on the length regime $\ell \ll L \ll \beta$,
where $\ell$ is the mean free path.

For the case of on-site disorder,
we see that as the disorder strength increases
($\sigma$ increases),
the Meissner stiffness decreases as expected.
The algebraically decaying part follows $L^{-2}$ and the exponents of the
\textcolor{black}{exponentially decaying part} grows as disorder strength is increased.
Such behavior fits the Anderson localization picture where the localization length decreases as the disorder strength increases. 
A similar behavior is also observed
for the random hopping model,
where the electrical Meissner stiffness also follows algebraically and
exponentially decay.
We note that 
the electrical conductance for the random hopping model
is known to decay algebraically, $g \sim 1/\sqrt{L}$.
The exponentially decaying part could be explained by the normalization of energy level spacing.

\subsection{Energy-twisted boundary condition}

\begin{figure*}[t]
	\centering
	\includegraphics[width = 0.32\linewidth]{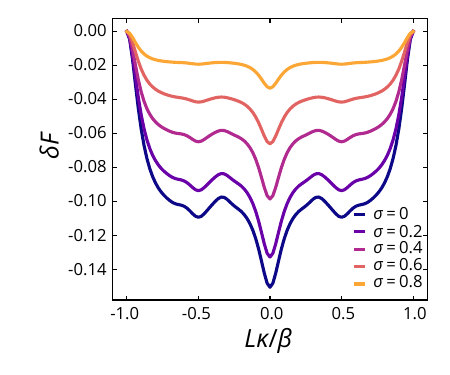}
  \includegraphics[width=0.32\linewidth]{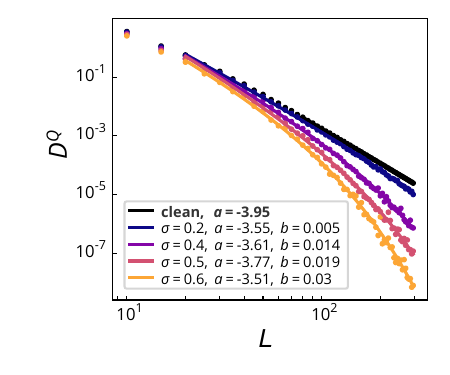}
  \includegraphics[width=0.32\linewidth]{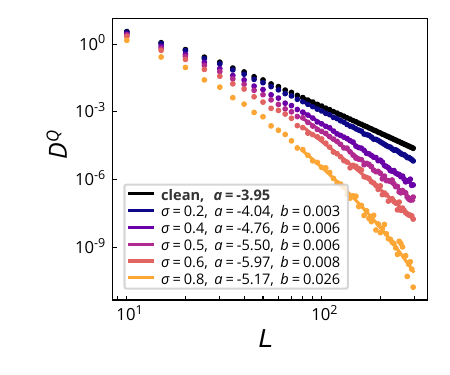}
	\caption{
    (Left)
 The energy-twist variation of the free energy of
    the free fermion model with random on-site potential.
        The curves are shown as varying $\sigma$ from $0$ to $0.8$. The calculation is done with $L = 20$ and $\beta = 100$. 
    The thermal Meissner stiffness for
    the disordered free fermion chain
    with
    random on-site potential (middle),
    and 
    with random hopping (right).
    Dots represent the results from the transfer matrix method
    with $\beta = 500$
    and solid lines are fitting with $D^Q \sim L^a e^{-bL}$.
		\label{fig:free_fermion_energy_twist}
  }
\end{figure*}

We now turn to the energy-twisted boundary condition.
\textcolor{black}{It can be} implemented in
the column-to-column transfer matrix method as 
\begin{equation}
	Z = \text{Tr}\, [T_{1,2}\cdots T_{N-1,N}e^{i P \Delta \tau}].
\end{equation}
Similar to \textcolor{black}{the} phase-twisted boundary condition,
the energy twist results in
the coupling $t_{i,i+1} \rightarrow t_{i,i+1}e^{i\omega M\kappa}$ where $\omega$ is
the frequency in the Trotter direction.
Therefore, following the same calculation as \textcolor{black}{the} phase twist,
the partition function can be written as 
\begin{equation}
Z = \prod_i C_i^M \prod_{\omega} \text{Tr}\, \big[2 \cos(\omega M \kappa)
+T_{\omega}\big].
\end{equation}

The energy-twisted free energy and the thermal
Meissner stiffness, computed by the transfer matrix method,
are plotted in Fig.\ \ref{fig:free_fermion_energy_twist}. 
As before, we study the clean fermion model,
\textcolor{black}{the} disordered model with on-site disorder,
and \textcolor{black}{the} random hopping model. For the free energy plot, we consider the system with on-site random potential $U_i$ to be Gaussian distributed with variance $\sigma$. 
\textcolor{black}{We could see that as the disorder strength increases, the free energy curves become more flat,} which means the system is more localized and less sensitive to boundary conditions. 

For the clean system,
we checked that the thermal Meissner stiffness decays
algebraically as
$D^{Q} \sim L^{-4}$
(for $L\ll \beta$),
which agrees with the CFT prediction $d^2 F/ d\kappa^2 \sim L^{-3}$. 
For the case of on-site disorder,
the 
\textcolor{black}{algebraically} decaying part generally follows $D^{Q} \sim L^{-4}$. As the
disorder is stronger, the thermal Meissner stiffness
decays exponentially with length as expected from Anderson localization.
\textcolor{black}{The exponent represents the inverse of the localization length and it increases as the disorder is stronger. }

For the random hopping model,
the thermal Meissner stiffness also shows algebraic and exponential decay as the case of Anderson localization. The electrical conductance of 
\textcolor{black}{the random hopping model} decays algebraically as $g \sim 1/\sqrt{L}$. Due to \textcolor{black}{the} Wiedemann-Franz law, we expect the thermal conductance also behaves similarly. The conductance is given by 
$g \sim {E}/{\Delta}$ with 
$E$ being the sensitivity of \textcolor{black}{the} energy to the twisted boundary condition and $\Delta = 1/\rho(0)$ is the energy level spacing at zero energy. The exponential decay might \textcolor{black}{be} due to the zero energy level spacing of \textcolor{black}{the} random hopping model.

\section{Boost deformation and thermal response}
\label{sec:thermaltransport_deformation}

In this section, we show that the thermal Drude weight and thermal Meissner stiffness of a disordered lattice fermion are related to the boost deformation via (\ref{eq:thermaldrudeweight_boostdeformation}) and (\ref{eq:thermalmeissnerstiffness_boostdeformation}).
The argument in this section is basically in parallel with the analogous
$U(1)$ twist.

\subsection{Thermal conductivity}
\label{sec:thermaltransport_deformation_thermalconductivity}

First, we review the thermal Drude weight $\bar{D}^Q$ and the thermal Meissner stiffness $D^Q$ following \cite{shastry06}.

The ac thermal conductivity of a local Hamiltonian $H_0=\sum_i H_i$ coupled with
a gravitational field $\psi_j(t)=e^{iqj-i(\omega+i\eta)t}$,
serving as a temperature profile via $\nabla\psi(r)=[\nabla T(r)]/T(r)$, is given in (\ref{eq:acthermalconductivity}) in the limit of $q\to 0$, where
\begin{align}
 &D^Q=
 \frac{1}{2L}\left(
 \langle\Theta\rangle-\int_0^\beta d\tau\langle J^Q(-i\tau)J^Q\rangle\right),\\
 &\kappa_\text{KG}(\omega)
 =
 -\frac{1}{LT}
 \int_{-\infty}^0 dt e^{-i(\omega+i\eta)t}
 \int_0^\beta d\tau \langle J^Q(t-i\tau)J^Q\rangle.
\end{align}
Here, $A(t)=e^{itH_0}Ae^{-itH_0}$, $\langle A\rangle=\text{Tr}[e^{-\beta H_0} A]/Z$, $Z=\text{Tr}[e^{-\beta H_0}]$ is the partition function, and the heat current and thermal operators are
\begin{align}
 &J^Q=\sum_jJ_j^Q=-\frac{i}{2}\sum_{jk}(j-k)[H_j,H_k],\\
 &\Theta=
 -\frac{1}{2}\sum_{jak}(j-k)(j-a)
 [[H_j,H_a],H_k].
\end{align}
Notice that these operators are defined unambiguously when the distance of two sites $j-k$ is uniquely defined, that is, when the Hamiltonian is local (the distance $|j-k|$ up to which $[H_j,H_k]\neq 0$ is bounded) or unless subject to PBC.

In terms of the eigenenergy $E_n$ and eigenstates $|n\rangle$ of the Hamiltonian $H_0$, the thermal Drude weight is
\begin{align}
 \bar{D}^Q
 &=
 D^Q
 +
 \frac{1}{2LT}\sum_{\substack{n,m\\E_n=E_m}}
 \frac{e^{-\beta E_n}}{Z}|\langle n|J^Q|m\rangle|^2 \notag\\
 &=\frac{1}{2L}\left(\langle\Theta\rangle-2\sum_{\substack{n,m\\E_n\neq E_m}}\frac{e^{-\beta E_n}}{Z}\frac{|\langle n|J^Q|m\rangle|^2}{E_m-E_n}\right).
 \label{eq:appendix_thermaldrudeweight}
\end{align}
In the limit of vanishing temperature ($T\to 0$) while keeping the system size finite ($L\ll \infty$), the thermal Drude weight and the thermal Meissner stiffness coincide unless the ground state is degenerate.

When a disordered, free lattice fermion Hamiltonian
\begin{align}
 H_0=\sum_{jk}^Lt_{jk}c_j^{\dagger}c_k^{\ }
\end{align}
is considered, the above operators are given, respectively, by
\begin{align}
 &J^Q=
 -\frac{i}{2}\sum_{jak}^L(j-k)t_{ja}t_{ak}c_j^{\dagger}c_k^{\ }, 
 \label{eq:appendix_heatcurrent_disorderedfermion}\\
 &\Theta=
 -\frac{1}{4}\sum_{jabk}^L
 (j-k)(j+a-b-k)t_{ja}t_{ab}t_{bk}c_j^{\dagger}c_k^{\ }
 \label{eq:appendix_thermaloperator_disorderedfermion}.
\end{align}

\subsection{Boost deformation}

We consider a disordered lattice fermion model and the corresponding boost operator given by
\begin{align}
 &H(\lambda)=\sum_{jk}t_{jk}(\lambda)c_j^{\dagger}c_k^{\ },
 \label{eq:appendix_disorderedhamiltonian}\\
 &\mathcal{B}[H(\lambda)]=
 \sum_{jk}\frac{j+k}{2}t_{jk}(\lambda)c_j^{\dagger}c_k^{\ }.
\end{align}
The boost deformation (\ref{eq:boostdeformation}) is reduced to 
\begin{align}
 \frac{dt_{jk}(\lambda)}{d\lambda}=
 \frac{i(j-k)}{2}\sum_a t_{ja}(\lambda)t_{ak}(\lambda),
\end{align}
and from this equation the second derivative is 
\begin{align}
 \frac{d^2t_{jk}(\lambda)}{d\lambda^2}=
 -\frac{j-k}{4}\sum_{ab}(j+a-b-k) t_{ja}(\lambda)t_{ab}(\lambda)t_{bk}(\lambda).
\end{align}
Notice that we adopted a specific Hamiltonian (\ref{eq:appendix_disorderedhamiltonian}) since the second derivative of a general local Hamiltonian $H=\sum H_j$ cannot be obtained in this way.
Referring to (\ref{eq:appendix_heatcurrent_disorderedfermion}) and (\ref{eq:appendix_thermaloperator_disorderedfermion}), the deformed Hamiltonian is expanded around $\lambda=0$ as
\begin{align}
 H(\lambda)
 =
 H(\lambda=0)-\lambda J^Q+\frac{\lambda^2}{2}\Theta + O(\lambda^3),
 \label{eq:appendix_hamiltonianexpansion_boostparameter}
\end{align}
where the operators $J^Q$ and $\Theta$ are defined with hopping parameters before the deformation $t_{jk}(\lambda=0)$.

From (\ref{eq:appendix_hamiltonianexpansion_boostparameter}), the perturbative expansion of an eigenenergy up to the second order in the boost parameter is
\begin{align}
 &E_{n}(\lambda)
 =
 E_n(\lambda=0)-\lambda\langle n|J^Q|n\rangle \notag\\
 &+
 \frac{\lambda^2}{2}\left(\langle n|\Theta|n\rangle-2\sum_{\substack{m\\E_n(0)\neq E_m(0)}}\frac{|\langle n|J^Q|m\rangle|^2}{E_m(0)- E_n(0)}\right),
\end{align}
which gives a relation between the thermal Drude weight (\ref{eq:appendix_thermaldrudeweight}) and the boost deformation as shown in (\ref{eq:thermaldrudeweight_boostdeformation}).

On the other hand, the derivative of the free energy $F(\lambda)=-\beta^{-1}\ln Z(\lambda)$ is, by using (\ref{eq:appendix_hamiltonianexpansion_boostparameter}) and the absence of the heat current $\langle J^Q\rangle=-dF/d\lambda|_{\lambda=0}$ in the ground state,
\begin{align}
 \frac{d^2F(\lambda)}{d\lambda^2}\bigg|_{\lambda=0}=
 \langle\Theta\rangle-
 \int_0^\beta d\tau \langle J^Q(-i\tau)J^Q\rangle,
\end{align}
which leads to the relation (\ref{eq:thermalmeissnerstiffness_boostdeformation}) between the thermal Meissner stiffness and the derivative of the free energy.

\subsection{A clean system in the thermodynamic limit}
\label{sec:appendix_thermaltransportcoefficients_cleanfermion}

We rederive the thermal Drude weight and thermal Meissner stiffness  of a clean lattice fermion in the thermodynamic limit \cite{shastry06} by using the boost deformation.
When spatial translation symmetry is present and in the thermodynamic limit $L\to \infty$, the single-particle eigenenergy $\epsilon_q=\sum_{a}t_{jj+a}e^{iqa}$ is a differentiable function of the momentum $q$ and the boost parameter $\lambda$, and thus the heat current and thermal operators are
\begin{align}
 &J^Q=-\sum_q \frac{\partial\epsilon_q}{\partial\lambda}c_q^{\dagger}c_q^{\ }=
 \sum_q \epsilon_q\frac{\partial\epsilon_q}{\partial q}c_q^{\dagger}c_q^{\ },\\
 &\Theta=\sum_q \frac{\partial^2\epsilon_q}{\partial\lambda^2}c_q^{\dagger}c_q^{\ }=
 \sum_q \frac{\partial}{\partial q}\left(\epsilon_q^2\frac{\partial\epsilon_q}{\partial q}\right)c_q^{\dagger}c_q^{\ },
\end{align}
where $c_q=L^{-1/2}\sum_j e^{iqj}c_j$.
The derivatives of the averaged many-body eigenenergy $E_n$ and the free energy $F=-\beta^{-1}\sum_q \ln(1+e^{-\beta\epsilon_q})$ are
\begin{align}
 &\sum_n\frac{e^{-\beta E_n}}{Z}\frac{d^2E_n}{d\lambda^2}=
 \sum_qf(\epsilon_q)\frac{\partial^2\epsilon_q}{\partial\lambda^2},\\
 &\frac{d^2F}{d\lambda^2}=
 \sum_q\left[f(\epsilon_q)\frac{\partial^2\epsilon_q}{\partial\lambda^2}+\frac{df(\epsilon_q)}{d\epsilon_q}\left(\frac{\partial\epsilon_q}{\partial\lambda}\right)^2\right],
\end{align}
where $f$ is the Fermi distribution function.
Substituting into (\ref{eq:thermaldrudeweight_boostdeformation}) and (\ref{eq:thermalmeissnerstiffness_boostdeformation}), we obtain
\begin{align}
 &\bar{D}^Q=
 \frac{1}{4\pi}\int dq \left(-\frac{df}{d\epsilon_q}\right)\left(\epsilon_q\frac{\partial\epsilon_q}{\partial q}\right)^2=\frac{\pi}{12\beta^2}\sum_\text{FP}|v_F|, 
 \\
 &D^Q=
 \frac{1}{4\pi}\int dq \frac{\partial}{\partial q}\left[f(\epsilon_q)\epsilon_q^2\frac{\partial\epsilon_q}{\partial q}\right]=0,
\end{align}
where FP stands for the Fermi points.

\subsection{Finite length behavior}
\label{sec:freefermion_boosttms}

In this subsection, we see that the thermal Meissner stiffness at $T=0$ depends qualitatively on the length modulo 4, and that they are related to the low-energy excitations.

\begin{figure}
 \includegraphics[width=70mm]{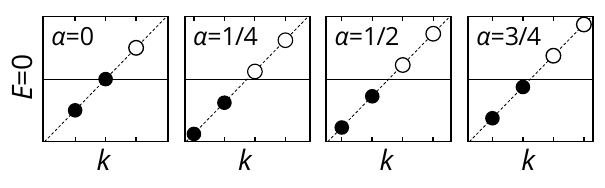}
 \includegraphics[width=70mm]{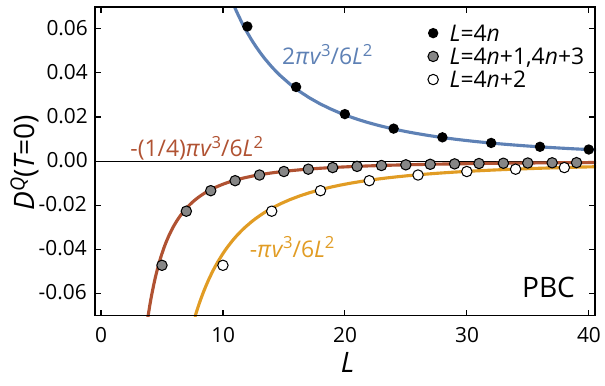}
 \includegraphics[width=70mm]{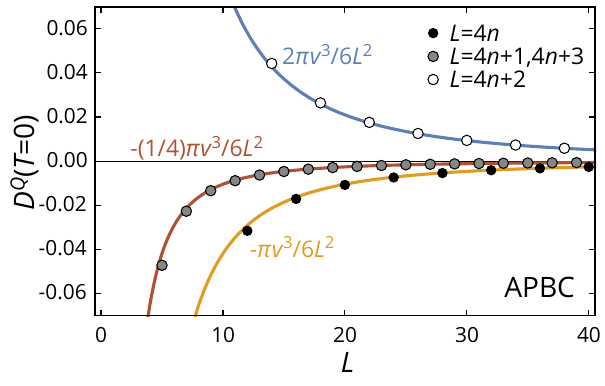}
 \caption{Low energy excitations for $\alpha=0,1/4,1/2$, and $3/4$.
 The thermal Meissner stiffness at zero temperature of a lattice free fermion is shown for PBC and APBC and is fitted by the corresponding CFT results. 
 \label{fig:tmsfreefermionmod4}}
\end{figure}

Figure \ref{fig:tmsfreefermionmod4} shows the detailed length dependence of the thermal Meissner stiffness at $T=0$ for PBC and APBC.
When the length is $L=4n(n\in \mathbb{N})$, the thermal Meissner stiffness of PBC scales 
as $2\pi v^3/6L^2$ while that of APBC 
scales as $-\pi v^3/6L^2$.
However, when the length is $L=4n+2$, 
these behaviors are inverted.
When the length is an odd integer ($L=4n+1$ or $4n+3$), the thermal Meissner stiffness 
scales as $-(1/4)\pi v^3/6L^2$.

As was shown in Sec.~\ref{sec:cft}, the thermal Meissner stiffness (\ref{eq:cft_thermalmeissnerstiffness}) of CFT at sufficiently low temperature is proportional to the ground state energy 
$
E_0=
-(2\pi v/L)
(c-24h)/12.
$
A one-dimensional Dirac fermion is equivalent to two real fermions corresponding to the Ising CFT, and hence the ground state energy of the Dirac fermion is equal to twice that of the Ising CFT ($c=1/2$).
Specifically, when a boundary condition $\psi(x+L)=e^{2\pi i\alpha}\psi(x)$ where $\alpha\in[0,1)$ is imposed, the single-particle eigenenergy of a chiral Dirac fermion $H=-iv\partial_x$ is $2\pi rv/L\,(r\in\mathbb{Z}+\alpha)$, and hence
\begin{align}
 H=\frac{2\pi v}{L}\sum_rrc_r^\dagger c_r^{\ }
 =\frac{2\pi v}{L}\sum_rr:c_r^\dagger c_r^{\ }:+E_0,
\end{align}
where $:\,\,:$ is the normal ordering, and via the zeta-function regularization,
\begin{align}
 E_0=\frac{2\pi v}{L}\sum_{n=1}^\infty(-n+\alpha)=-\frac{2\pi v}{L}\left[\frac{1}{24}-\frac{1}{2}\left(\frac{1}{2}-\alpha\right)^2\right].
 \label{eq:groundstateenergy_zeta}
\end{align}
As for the left mover, we impose $\psi(x+L)=e^{-2\pi i\alpha}\psi(x)$ to make $\alpha$-dependence of the energy levels the same as the right one.
Then the ground state energy of the helical Dirac fermion 
(including both left and right movers)
with PBC ($\alpha=0$) is $E_0=(1/6)(2\pi v/L)$ that corresponds to twice the ground state energy of CFT with $c=1/2$ and $h=1/16$, and that with APBC ($\alpha=1/2$) is $E_0=(-1/12)(2\pi v/L)$ corresponding to $c=1/2$ and $h=0$.

To make a connection to the lattice fermion, we naively anticipate that the ground state energy used for deriving the thermal Meissner stiffness can be identified with that of the linearized helical Dirac fermion, since low-energy states are relevant to low-temperature behavior.
\textcolor{black}{In doing so, we notice that energy levels near the Fermi level depend on the length} and the boundary condition (Fig.~\ref{fig:tmsfreefermionmod4}).
Let us assume the cosine band $\epsilon_k=-2\cos k$. With PBC and $L=4n$, there are states $k_F=\pm\pi/2$ exactly at the Fermi level and thus the right and left movers correspond to $\alpha=0$. 
Similarly, PBC with $L=4n+2$ corresponds to $\alpha=1/2$, and PBC with $L=4n+1(4n+3)$ to $\alpha=3/4(1/4)$.
Specifically, when the length is odd, the ground state energy is $-(1/48)2\pi v/L$ from (\ref{eq:groundstateenergy_zeta}), and the corresponding thermal Meissner stiffness is estimated as $D^Q=-(1/4)(\pi v^3/6L^2)$.
Strictly speaking, the ground state energy obtained in this way is not the actual energy, but a quantity related to thermal response.
When switched to APBC, the above results still hold by shifting $\alpha\to\alpha+1/2 \text{ mod } 1$, and hence this explains mod 4 behavior seen in Fig.~\ref{fig:tmsfreefermionmod4}.
Notice that this argument is true when the chemical potential is 0, where $\alpha$ depends on the length only modulo 4.

\section{Quantum Hall systems with boost deformation}
\label{Quantum Hall system with boost deformation}

In this Appendix,
we consider the boost deformation
of the quantum Hall system.
We start with the Hamiltonian
of 2d electron gas in the presence of
uniform magnetic field, 
\begin{equation}
    H = \frac{1}{2m} (-i\hbar \bold{\partial} - e\bold{A})^2,
\end{equation}
with Landau gauge $\bold{A} = (-By,0)$. 
We consider the cylinder geometry
with periodic $x$ direction.
The energy levels (Landau level)
are given by
$
    \varepsilon_N = \hbar \omega_c (N+\frac{1}{2})
$
where $\omega_c = |e|B/m$.
The corresponding 
wave functions for the $N$-th Landau level
are given by
\begin{equation}
    \psi_{N,p_x}(x,y) \propto e^{ip_x x}e^{-(y-y_0)^2/2l^2}H_N(y-y_0),
\end{equation}
where $y_0 = \hbar p_x/eB$, $l$ is the magnetic length,
and $H_N$ is the Hermite polynomial. 

If we consider the boost deformation in $x$ direction
and impose 
the energy-twisted boundary condition,
this amounts to shifting single-particle momentum,
$p_x \rightarrow p_x+ \lambda \varepsilon_N$.
From the periodicity in $x$ direction,
$(p_x + \lambda \varepsilon_N)L = 2\pi r$
where $r$ is an integer. 
As $\varepsilon_N$ does not depend on $p_x$,
this equation can be readily solved, 
\begin{align}
  p_x =
  \frac{2\pi}{L}
  \left( r -
  \frac{L \lambda \varepsilon_N}{2\pi}
  \right).
\end{align}
As we change $\lambda$
from $0$ to $ (2\pi)/(L\varepsilon_N)$, 
$p_x$ changes from $2\pi r/L$ to $2\pi (r-1)/L$.
This results in the shift of the Landau level center,
$y_0 =
(\hbar p_x)/(e B)
=
(\hbar 2\pi r)/ (e B L)
\to
(\hbar 2\pi (r+1))/ (e B L)
$.
As Laughlin's argument for the quantized
Hall conductance, 
the adiabatic change in $\lambda$
transports
one electron
($\equiv \Delta N$)
from
one end of the cylinder to the other.
Hence,
$
\Delta N/\Delta \lambda
=
1/[( 2\pi) / (L\varepsilon_N)]
=
(L \varepsilon_N)/(2\pi)
$.
The transported energy
$\Delta E = \varepsilon_N \Delta N$
is given by
$
(\Delta E)/(\Delta \lambda)
=
(L \varepsilon^2_N)/(2\pi)
$.
Following the 
analogy to 
Laughlin's argument for the quantized
Hall conductance, 
the transverse energy transport
is induced by
the adiabatic insertion of 
a boost-analogue of magnetic flux.

\bibliography{reference}

\begin{thebibliography}{48}%
\makeatletter
\providecommand \@ifxundefined [1]{%
 \@ifx{#1\undefined}
}%
\providecommand \@ifnum [1]{%
 \ifnum #1\expandafter \@firstoftwo
 \else \expandafter \@secondoftwo
 \fi
}%
\providecommand \@ifx [1]{%
 \ifx #1\expandafter \@firstoftwo
 \else \expandafter \@secondoftwo
 \fi
}%
\providecommand \natexlab [1]{#1}%
\providecommand \enquote  [1]{``#1''}%
\providecommand \bibnamefont  [1]{#1}%
\providecommand \bibfnamefont [1]{#1}%
\providecommand \citenamefont [1]{#1}%
\providecommand \href@noop [0]{\@secondoftwo}%
\providecommand \href [0]{\begingroup \@sanitize@url \@href}%
\providecommand \@href[1]{\@@startlink{#1}\@@href}%
\providecommand \@@href[1]{\endgroup#1\@@endlink}%
\providecommand \@sanitize@url [0]{\catcode `\\12\catcode `\$12\catcode
  `\&12\catcode `\#12\catcode `\^12\catcode `\_12\catcode `\%12\relax}%
\providecommand \@@startlink[1]{}%
\providecommand \@@endlink[0]{}%
\providecommand \url  [0]{\begingroup\@sanitize@url \@url }%
\providecommand \@url [1]{\endgroup\@href {#1}{\urlprefix }}%
\providecommand \urlprefix  [0]{URL }%
\providecommand \Eprint [0]{\href }%
\providecommand \doibase [0]{https://doi.org/}%
\providecommand \selectlanguage [0]{\@gobble}%
\providecommand \bibinfo  [0]{\@secondoftwo}%
\providecommand \bibfield  [0]{\@secondoftwo}%
\providecommand \translation [1]{[#1]}%
\providecommand \BibitemOpen [0]{}%
\providecommand \bibitemStop [0]{}%
\providecommand \bibitemNoStop [0]{.\EOS\space}%
\providecommand \EOS [0]{\spacefactor3000\relax}%
\providecommand \BibitemShut  [1]{\csname bibitem#1\endcsname}%
\let\auto@bib@innerbib\@empty
\bibitem [{\citenamefont {Luttinger}(1964)}]{luttinger64}%
  \BibitemOpen
  \bibfield  {author} {\bibinfo {author} {\bibfnamefont {J.~M.}\ \bibnamefont
  {Luttinger}},\ }\bibfield  {title} {\bibinfo {title} {Theory of thermal
  transport coefficients},\ }\href {https://doi.org/10.1103/PhysRev.135.A1505}
  {\bibfield  {journal} {\bibinfo  {journal} {Phys. Rev.}\ }\textbf {\bibinfo
  {volume} {135}},\ \bibinfo {pages} {A1505} (\bibinfo {year}
  {1964})}\BibitemShut {NoStop}%
\bibitem [{\citenamefont {{Kohn}}(1964)}]{1964PhRv..133..171K}%
  \BibitemOpen
  \bibfield  {author} {\bibinfo {author} {\bibfnamefont {W.}~\bibnamefont
  {{Kohn}}},\ }\bibfield  {title} {\bibinfo {title} {{Theory of the Insulating
  State}},\ }\href {https://doi.org/10.1103/PhysRev.133.A171} {\bibfield
  {journal} {\bibinfo  {journal} {Physical Review}\ }\textbf {\bibinfo {volume}
  {133}},\ \bibinfo {pages} {171} (\bibinfo {year} {1964})}\BibitemShut
  {NoStop}%
\bibitem [{\citenamefont {{Edwards}}\ and\ \citenamefont
  {{Thouless}}(1972)}]{1972JPhC....5..807E}%
  \BibitemOpen
  \bibfield  {author} {\bibinfo {author} {\bibfnamefont {J.~T.}\ \bibnamefont
  {{Edwards}}}\ and\ \bibinfo {author} {\bibfnamefont {D.~J.}\ \bibnamefont
  {{Thouless}}},\ }\bibfield  {title} {\bibinfo {title} {{Numerical studies of
  localization in disordered systems}},\ }\href
  {https://doi.org/10.1088/0022-3719/5/8/007} {\bibfield  {journal} {\bibinfo
  {journal} {Journal of Physics C Solid State Physics}\ }\textbf {\bibinfo
  {volume} {5}},\ \bibinfo {pages} {807} (\bibinfo {year} {1972})}\BibitemShut
  {NoStop}%
\bibitem [{\citenamefont {Thouless}(1977)}]{PhysRevLett.39.1167}%
  \BibitemOpen
  \bibfield  {author} {\bibinfo {author} {\bibfnamefont {D.~J.}\ \bibnamefont
  {Thouless}},\ }\bibfield  {title} {\bibinfo {title} {Maximum metallic
  resistance in thin wires},\ }\href
  {https://doi.org/10.1103/PhysRevLett.39.1167} {\bibfield  {journal} {\bibinfo
   {journal} {Phys. Rev. Lett.}\ }\textbf {\bibinfo {volume} {39}},\ \bibinfo
  {pages} {1167} (\bibinfo {year} {1977})}\BibitemShut {NoStop}%
\bibitem [{\citenamefont {Nakai}\ \emph {et~al.}(2017)\citenamefont {Nakai},
  \citenamefont {Ryu},\ and\ \citenamefont {Nomura}}]{PhysRevB.95.165405}%
  \BibitemOpen
  \bibfield  {author} {\bibinfo {author} {\bibfnamefont {R.}~\bibnamefont
  {Nakai}}, \bibinfo {author} {\bibfnamefont {S.}~\bibnamefont {Ryu}},\ and\
  \bibinfo {author} {\bibfnamefont {K.}~\bibnamefont {Nomura}},\ }\bibfield
  {title} {\bibinfo {title} {Laughlin's argument for the quantized thermal hall
  effect},\ }\href {https://doi.org/10.1103/PhysRevB.95.165405} {\bibfield
  {journal} {\bibinfo  {journal} {Phys. Rev. B}\ }\textbf {\bibinfo {volume}
  {95}},\ \bibinfo {pages} {165405} (\bibinfo {year} {2017})}\BibitemShut
  {NoStop}%
\bibitem [{\citenamefont {{Golkar}}\ and\ \citenamefont
  {{Sethi}}(2015)}]{2015arXiv151202607G}%
  \BibitemOpen
  \bibfield  {author} {\bibinfo {author} {\bibfnamefont {S.}~\bibnamefont
  {{Golkar}}}\ and\ \bibinfo {author} {\bibfnamefont {S.}~\bibnamefont
  {{Sethi}}},\ }\bibfield  {title} {\bibinfo {title} {{Global Anomalies and
  Effective Field Theory}},\ }\href@noop {} {\bibfield  {journal} {\bibinfo
  {journal} {arXiv e-prints}\ ,\ \bibinfo {eid} {arXiv:1512.02607}} (\bibinfo
  {year} {2015})},\ \Eprint {https://arxiv.org/abs/1512.02607}
  {arXiv:1512.02607 [hep-th]} \BibitemShut {NoStop}%
\bibitem [{\citenamefont {{B{\"u}ttiker}}\ \emph {et~al.}(1983)\citenamefont
  {{B{\"u}ttiker}}, \citenamefont {{Imry}},\ and\ \citenamefont
  {{Landauer}}}]{1983PhLA...96..365B}%
  \BibitemOpen
  \bibfield  {author} {\bibinfo {author} {\bibfnamefont {M.}~\bibnamefont
  {{B{\"u}ttiker}}}, \bibinfo {author} {\bibfnamefont {Y.}~\bibnamefont
  {{Imry}}},\ and\ \bibinfo {author} {\bibfnamefont {R.}~\bibnamefont
  {{Landauer}}},\ }\bibfield  {title} {\bibinfo {title} {{Josephson behavior in
  small normal one-dimensional rings}},\ }\href
  {https://doi.org/10.1016/0375-9601(83)90011-7} {\bibfield  {journal}
  {\bibinfo  {journal} {Physics Letters A}\ }\textbf {\bibinfo {volume} {96}},\
  \bibinfo {pages} {365} (\bibinfo {year} {1983})}\BibitemShut {NoStop}%
\bibitem [{\citenamefont {Rizzi}\ and\ \citenamefont
  {Ruggiero}(2003)}]{Rizzi2003-ka}%
  \BibitemOpen
  \bibinfo {editor} {\bibfnamefont {G.}~\bibnamefont {Rizzi}}\ and\ \bibinfo
  {editor} {\bibfnamefont {M.~L.}\ \bibnamefont {Ruggiero}},\ eds.,\ \href@noop
  {} {\emph {\bibinfo {title} {Relativity in rotating frames}}},\ \bibinfo
  {edition} {2004th}\ ed.,\ Fundamental Theories of Physics\ (\bibinfo
  {publisher} {Springer},\ \bibinfo {address} {New York, NY},\ \bibinfo {year}
  {2003})\BibitemShut {NoStop}%
\bibitem [{\citenamefont {{Shiozaki}}\ \emph {et~al.}(2018)\citenamefont
  {{Shiozaki}}, \citenamefont {{Shapourian}}, \citenamefont {{Gomi}},\ and\
  \citenamefont {{Ryu}}}]{2018PhRvB..98c5151S}%
  \BibitemOpen
  \bibfield  {author} {\bibinfo {author} {\bibfnamefont {K.}~\bibnamefont
  {{Shiozaki}}}, \bibinfo {author} {\bibfnamefont {H.}~\bibnamefont
  {{Shapourian}}}, \bibinfo {author} {\bibfnamefont {K.}~\bibnamefont
  {{Gomi}}},\ and\ \bibinfo {author} {\bibfnamefont {S.}~\bibnamefont
  {{Ryu}}},\ }\bibfield  {title} {\bibinfo {title} {{Many-body topological
  invariants for fermionic short-range entangled topological phases protected
  by antiunitary symmetries}},\ }\href
  {https://doi.org/10.1103/PhysRevB.98.035151} {\bibfield  {journal} {\bibinfo
  {journal} {\prb}\ }\textbf {\bibinfo {volume} {98}},\ \bibinfo {eid} {035151}
  (\bibinfo {year} {2018})},\ \Eprint {https://arxiv.org/abs/1710.01886}
  {arXiv:1710.01886 [cond-mat.str-el]} \BibitemShut {NoStop}%
\bibitem [{\citenamefont {{Shapourian}}\ \emph {et~al.}(2017)\citenamefont
  {{Shapourian}}, \citenamefont {{Shiozaki}},\ and\ \citenamefont
  {{Ryu}}}]{2017PhRvL.118u6402S}%
  \BibitemOpen
  \bibfield  {author} {\bibinfo {author} {\bibfnamefont {H.}~\bibnamefont
  {{Shapourian}}}, \bibinfo {author} {\bibfnamefont {K.}~\bibnamefont
  {{Shiozaki}}},\ and\ \bibinfo {author} {\bibfnamefont {S.}~\bibnamefont
  {{Ryu}}},\ }\bibfield  {title} {\bibinfo {title} {{Many-Body Topological
  Invariants for Fermionic Symmetry-Protected Topological Phases}},\ }\href
  {https://doi.org/10.1103/PhysRevLett.118.216402} {\bibfield  {journal}
  {\bibinfo  {journal} {\prl}\ }\textbf {\bibinfo {volume} {118}},\ \bibinfo
  {eid} {216402} (\bibinfo {year} {2017})},\ \Eprint
  {https://arxiv.org/abs/1607.03896} {arXiv:1607.03896 [cond-mat.str-el]}
  \BibitemShut {NoStop}%
\bibitem [{\citenamefont {Trivedi}\ and\ \citenamefont
  {Browne}(1988)}]{trivedi88}%
  \BibitemOpen
  \bibfield  {author} {\bibinfo {author} {\bibfnamefont {N.}~\bibnamefont
  {Trivedi}}\ and\ \bibinfo {author} {\bibfnamefont {D.~A.}\ \bibnamefont
  {Browne}},\ }\bibfield  {title} {\bibinfo {title} {Mesoscopic ring in a
  magnetic field: Reactive and dissipative response},\ }\href
  {https://doi.org/10.1103/PhysRevB.38.9581} {\bibfield  {journal} {\bibinfo
  {journal} {Phys. Rev. B}\ }\textbf {\bibinfo {volume} {38}},\ \bibinfo
  {pages} {9581} (\bibinfo {year} {1988})}\BibitemShut {NoStop}%
\bibitem [{\citenamefont {Scalapino}\ \emph {et~al.}(1993)\citenamefont
  {Scalapino}, \citenamefont {White},\ and\ \citenamefont
  {Zhang}}]{PhysRevB.47.7995}%
  \BibitemOpen
  \bibfield  {author} {\bibinfo {author} {\bibfnamefont {D.~J.}\ \bibnamefont
  {Scalapino}}, \bibinfo {author} {\bibfnamefont {S.~R.}\ \bibnamefont
  {White}},\ and\ \bibinfo {author} {\bibfnamefont {S.}~\bibnamefont {Zhang}},\
  }\bibfield  {title} {\bibinfo {title} {Insulator, metal, or superconductor:
  The criteria},\ }\href {https://doi.org/10.1103/PhysRevB.47.7995} {\bibfield
  {journal} {\bibinfo  {journal} {Phys. Rev. B}\ }\textbf {\bibinfo {volume}
  {47}},\ \bibinfo {pages} {7995} (\bibinfo {year} {1993})}\BibitemShut
  {NoStop}%
\bibitem [{\citenamefont {Giamarchi}\ and\ \citenamefont
  {Shastry}(1995)}]{giamarchi95}%
  \BibitemOpen
  \bibfield  {author} {\bibinfo {author} {\bibfnamefont {T.}~\bibnamefont
  {Giamarchi}}\ and\ \bibinfo {author} {\bibfnamefont {B.~S.}\ \bibnamefont
  {Shastry}},\ }\bibfield  {title} {\bibinfo {title} {Persistent currents in a
  one-dimensional ring for a disordered hubbard model},\ }\href
  {https://doi.org/10.1103/PhysRevB.51.10915} {\bibfield  {journal} {\bibinfo
  {journal} {Phys. Rev. B}\ }\textbf {\bibinfo {volume} {51}},\ \bibinfo
  {pages} {10915} (\bibinfo {year} {1995})}\BibitemShut {NoStop}%
\bibitem [{\citenamefont {Shastry}(2006)}]{shastry06}%
  \BibitemOpen
  \bibfield  {author} {\bibinfo {author} {\bibfnamefont {B.~S.}\ \bibnamefont
  {Shastry}},\ }\bibfield  {title} {\bibinfo {title} {Sum rule for thermal
  conductivity and dynamical thermal transport coefficients in condensed
  matter},\ }\href {https://doi.org/10.1103/PhysRevB.73.085117} {\bibfield
  {journal} {\bibinfo  {journal} {Phys. Rev. B}\ }\textbf {\bibinfo {volume}
  {73}},\ \bibinfo {pages} {085117} (\bibinfo {year} {2006})}\BibitemShut
  {NoStop}%
\bibitem [{\citenamefont {Resta}(2018)}]{Resta_2018}%
  \BibitemOpen
  \bibfield  {author} {\bibinfo {author} {\bibfnamefont {R.}~\bibnamefont
  {Resta}},\ }\bibfield  {title} {\bibinfo {title} {Drude weight and
  superconducting weight},\ }\href {https://doi.org/10.1088/1361-648x/aade19}
  {\bibfield  {journal} {\bibinfo  {journal} {Journal of Physics: Condensed
  Matter}\ }\textbf {\bibinfo {volume} {30}},\ \bibinfo {pages} {414001}
  (\bibinfo {year} {2018})}\BibitemShut {NoStop}%
\bibitem [{\citenamefont {Castella}\ \emph {et~al.}(1995)\citenamefont
  {Castella}, \citenamefont {Zotos},\ and\ \citenamefont
  {Prelov\ifmmode~\check{s}\else \v{s}\fi{}ek}}]{PhysRevLett.74.972}%
  \BibitemOpen
  \bibfield  {author} {\bibinfo {author} {\bibfnamefont {H.}~\bibnamefont
  {Castella}}, \bibinfo {author} {\bibfnamefont {X.}~\bibnamefont {Zotos}},\
  and\ \bibinfo {author} {\bibfnamefont {P.}~\bibnamefont
  {Prelov\ifmmode~\check{s}\else \v{s}\fi{}ek}},\ }\bibfield  {title} {\bibinfo
  {title} {Integrability and ideal conductance at finite temperatures},\ }\href
  {https://doi.org/10.1103/PhysRevLett.74.972} {\bibfield  {journal} {\bibinfo
  {journal} {Phys. Rev. Lett.}\ }\textbf {\bibinfo {volume} {74}},\ \bibinfo
  {pages} {972} (\bibinfo {year} {1995})}\BibitemShut {NoStop}%
\bibitem [{\citenamefont {Zotos}\ \emph {et~al.}(1997)\citenamefont {Zotos},
  \citenamefont {Naef},\ and\ \citenamefont {Prelovsek}}]{zotos97}%
  \BibitemOpen
  \bibfield  {author} {\bibinfo {author} {\bibfnamefont {X.}~\bibnamefont
  {Zotos}}, \bibinfo {author} {\bibfnamefont {F.}~\bibnamefont {Naef}},\ and\
  \bibinfo {author} {\bibfnamefont {P.}~\bibnamefont {Prelovsek}},\ }\bibfield
  {title} {\bibinfo {title} {Transport and conservation laws},\ }\href
  {https://doi.org/10.1103/PhysRevB.55.11029} {\bibfield  {journal} {\bibinfo
  {journal} {Phys. Rev. B}\ }\textbf {\bibinfo {volume} {55}},\ \bibinfo
  {pages} {11029} (\bibinfo {year} {1997})}\BibitemShut {NoStop}%
\bibitem [{\citenamefont {Fujimoto}\ and\ \citenamefont
  {Kawakami}(1998)}]{Fujimoto_1998}%
  \BibitemOpen
  \bibfield  {author} {\bibinfo {author} {\bibfnamefont {S.}~\bibnamefont
  {Fujimoto}}\ and\ \bibinfo {author} {\bibfnamefont {N.}~\bibnamefont
  {Kawakami}},\ }\bibfield  {title} {\bibinfo {title} {Exact drude weight for
  the one-dimensional hubbard model at finite temperatures},\ }\href
  {https://doi.org/10.1088/0305-4470/31/2/008} {\bibfield  {journal} {\bibinfo
  {journal} {Journal of Physics A: Mathematical and General}\ }\textbf
  {\bibinfo {volume} {31}},\ \bibinfo {pages} {465} (\bibinfo {year}
  {1998})}\BibitemShut {NoStop}%
\bibitem [{\citenamefont {Mukerjee}\ and\ \citenamefont
  {Shastry}(2008)}]{PhysRevB.77.245131}%
  \BibitemOpen
  \bibfield  {author} {\bibinfo {author} {\bibfnamefont {S.}~\bibnamefont
  {Mukerjee}}\ and\ \bibinfo {author} {\bibfnamefont {B.~S.}\ \bibnamefont
  {Shastry}},\ }\bibfield  {title} {\bibinfo {title} {Signatures of diffusion
  and ballistic transport in the stiffness, dynamical correlation functions,
  and statistics of one-dimensional systems},\ }\href
  {https://doi.org/10.1103/PhysRevB.77.245131} {\bibfield  {journal} {\bibinfo
  {journal} {Phys. Rev. B}\ }\textbf {\bibinfo {volume} {77}},\ \bibinfo
  {pages} {245131} (\bibinfo {year} {2008})}\BibitemShut {NoStop}%
\bibitem [{\citenamefont {Tu}\ \emph {et~al.}(2013)\citenamefont {Tu},
  \citenamefont {Zhang},\ and\ \citenamefont {Qi}}]{PhysRevB.88.195412}%
  \BibitemOpen
  \bibfield  {author} {\bibinfo {author} {\bibfnamefont {H.-H.}\ \bibnamefont
  {Tu}}, \bibinfo {author} {\bibfnamefont {Y.}~\bibnamefont {Zhang}},\ and\
  \bibinfo {author} {\bibfnamefont {X.-L.}\ \bibnamefont {Qi}},\ }\bibfield
  {title} {\bibinfo {title} {Momentum polarization: An entanglement measure of
  topological spin and chiral central charge},\ }\href
  {https://doi.org/10.1103/PhysRevB.88.195412} {\bibfield  {journal} {\bibinfo
  {journal} {Phys. Rev. B}\ }\textbf {\bibinfo {volume} {88}},\ \bibinfo
  {pages} {195412} (\bibinfo {year} {2013})}\BibitemShut {NoStop}%
\bibitem [{\citenamefont {You}\ and\ \citenamefont {Cheng}(2015)}]{you15}%
  \BibitemOpen
  \bibfield  {author} {\bibinfo {author} {\bibfnamefont {Y.-Z.}\ \bibnamefont
  {You}}\ and\ \bibinfo {author} {\bibfnamefont {M.}~\bibnamefont {Cheng}},\
  }\bibfield  {title} {\bibinfo {title} {{Measuring Modular Matrices by
  Shearing Lattices}},\ }\href@noop {} {\  (\bibinfo {year} {2015})},\ \Eprint
  {https://arxiv.org/abs/1502.03192} {arXiv:1502.03192 [cond-mat.str-el]}
  \BibitemShut {NoStop}%
\bibitem [{\citenamefont {{Yao}}\ and\ \citenamefont
  {{Oshikawa}}(2021)}]{2021PhRvL.126u7201Y}%
  \BibitemOpen
  \bibfield  {author} {\bibinfo {author} {\bibfnamefont {Y.}~\bibnamefont
  {{Yao}}}\ and\ \bibinfo {author} {\bibfnamefont {M.}~\bibnamefont
  {{Oshikawa}}},\ }\bibfield  {title} {\bibinfo {title} {{Twisted Boundary
  Condition and Lieb-Schultz-Mattis Ingappability for Discrete Symmetries}},\
  }\href {https://doi.org/10.1103/PhysRevLett.126.217201} {\bibfield  {journal}
  {\bibinfo  {journal} {\prl}\ }\textbf {\bibinfo {volume} {126}},\ \bibinfo
  {eid} {217201} (\bibinfo {year} {2021})},\ \Eprint
  {https://arxiv.org/abs/2010.09244} {arXiv:2010.09244 [cond-mat.str-el]}
  \BibitemShut {NoStop}%
\bibitem [{\citenamefont {{Aksoy}}\ \emph {et~al.}(2021)\citenamefont
  {{Aksoy}}, \citenamefont {{Tiwari}},\ and\ \citenamefont
  {{Mudry}}}]{2021PhRvB.104g5146A}%
  \BibitemOpen
  \bibfield  {author} {\bibinfo {author} {\bibfnamefont {{\"O}.~M.}\
  \bibnamefont {{Aksoy}}}, \bibinfo {author} {\bibfnamefont {A.}~\bibnamefont
  {{Tiwari}}},\ and\ \bibinfo {author} {\bibfnamefont {C.}~\bibnamefont
  {{Mudry}}},\ }\bibfield  {title} {\bibinfo {title} {{Lieb-Schultz-Mattis type
  theorems for Majorana models with discrete symmetries}},\ }\href
  {https://doi.org/10.1103/PhysRevB.104.075146} {\bibfield  {journal} {\bibinfo
   {journal} {\prb}\ }\textbf {\bibinfo {volume} {104}},\ \bibinfo {eid}
  {075146} (\bibinfo {year} {2021})},\ \Eprint
  {https://arxiv.org/abs/2102.08389} {arXiv:2102.08389 [cond-mat.str-el]}
  \BibitemShut {NoStop}%
\bibitem [{\citenamefont {{Bargheer}}\ \emph {et~al.}(2009)\citenamefont
  {{Bargheer}}, \citenamefont {{Beisert}},\ and\ \citenamefont
  {{Loebbert}}}]{2009JPhA...42B5205B}%
  \BibitemOpen
  \bibfield  {author} {\bibinfo {author} {\bibfnamefont {T.}~\bibnamefont
  {{Bargheer}}}, \bibinfo {author} {\bibfnamefont {N.}~\bibnamefont
  {{Beisert}}},\ and\ \bibinfo {author} {\bibfnamefont {F.}~\bibnamefont
  {{Loebbert}}},\ }\bibfield  {title} {\bibinfo {title} {{Long-range
  deformations for integrable spin chains}},\ }\href
  {https://doi.org/10.1088/1751-8113/42/28/285205} {\bibfield  {journal}
  {\bibinfo  {journal} {Journal of Physics A Mathematical General}\ }\textbf
  {\bibinfo {volume} {42}},\ \bibinfo {eid} {285205} (\bibinfo {year}
  {2009})},\ \Eprint {https://arxiv.org/abs/0902.0956} {arXiv:0902.0956
  [hep-th]} \BibitemShut {NoStop}%
\bibitem [{\citenamefont {{Zamolodchikov}}(2004)}]{2004hep.th....1146Z}%
  \BibitemOpen
  \bibfield  {author} {\bibinfo {author} {\bibfnamefont {A.~B.}\ \bibnamefont
  {{Zamolodchikov}}},\ }\bibfield  {title} {\bibinfo {title} {{Expectation
  value of composite field $T{\bar T}$ in two-dimensional quantum field
  theory}},\ }\href@noop {} {\bibfield  {journal} {\bibinfo  {journal} {arXiv
  e-prints}\ ,\ \bibinfo {eid} {hep-th/0401146}} (\bibinfo {year} {2004})},\
  \Eprint {https://arxiv.org/abs/hep-th/0401146} {arXiv:hep-th/0401146
  [hep-th]} \BibitemShut {NoStop}%
\bibitem [{\citenamefont {{Pozsgay}}(2020)}]{2020ScPP....8...16P}%
  \BibitemOpen
  \bibfield  {author} {\bibinfo {author} {\bibfnamefont {B.}~\bibnamefont
  {{Pozsgay}}},\ }\bibfield  {title} {\bibinfo {title} {{Current operators in
  integrable spin chains: lessons from long range deformations}},\ }\href
  {https://doi.org/10.21468/SciPostPhys.8.2.016} {\bibfield  {journal}
  {\bibinfo  {journal} {SciPost Physics}\ }\textbf {\bibinfo {volume} {8}},\
  \bibinfo {eid} {016} (\bibinfo {year} {2020})},\ \Eprint
  {https://arxiv.org/abs/1910.12833} {arXiv:1910.12833 [cond-mat.stat-mech]}
  \BibitemShut {NoStop}%
\bibitem [{\citenamefont {Klümper}\ and\ \citenamefont
  {Sakai}(2002)}]{Kl_mper_2002}%
  \BibitemOpen
  \bibfield  {author} {\bibinfo {author} {\bibfnamefont {A.}~\bibnamefont
  {Klümper}}\ and\ \bibinfo {author} {\bibfnamefont {K.}~\bibnamefont
  {Sakai}},\ }\bibfield  {title} {\bibinfo {title} {The thermal conductivity of
  the spin-{\textonehalf} {XXZ} chain at arbitrary temperature},\ }\href
  {https://doi.org/10.1088/0305-4470/35/9/307} {\bibfield  {journal} {\bibinfo
  {journal} {Journal of Physics A: Mathematical and General}\ }\textbf
  {\bibinfo {volume} {35}},\ \bibinfo {pages} {2173} (\bibinfo {year}
  {2002})}\BibitemShut {NoStop}%
\bibitem [{\citenamefont {Alvarez}\ and\ \citenamefont
  {Gros}(2002)}]{PhysRevLett.89.156603}%
  \BibitemOpen
  \bibfield  {author} {\bibinfo {author} {\bibfnamefont {J.~V.}\ \bibnamefont
  {Alvarez}}\ and\ \bibinfo {author} {\bibfnamefont {C.}~\bibnamefont {Gros}},\
  }\bibfield  {title} {\bibinfo {title} {Anomalous thermal conductivity of
  frustrated heisenberg spin chains and ladders},\ }\href
  {https://doi.org/10.1103/PhysRevLett.89.156603} {\bibfield  {journal}
  {\bibinfo  {journal} {Phys. Rev. Lett.}\ }\textbf {\bibinfo {volume} {89}},\
  \bibinfo {pages} {156603} (\bibinfo {year} {2002})}\BibitemShut {NoStop}%
\bibitem [{\citenamefont {Heidrich-Meisner}\ \emph {et~al.}(2002)\citenamefont
  {Heidrich-Meisner}, \citenamefont {Honecker}, \citenamefont {Cabra},\ and\
  \citenamefont {Brenig}}]{PhysRevB.66.140406}%
  \BibitemOpen
  \bibfield  {author} {\bibinfo {author} {\bibfnamefont {F.}~\bibnamefont
  {Heidrich-Meisner}}, \bibinfo {author} {\bibfnamefont {A.}~\bibnamefont
  {Honecker}}, \bibinfo {author} {\bibfnamefont {D.~C.}\ \bibnamefont
  {Cabra}},\ and\ \bibinfo {author} {\bibfnamefont {W.}~\bibnamefont
  {Brenig}},\ }\bibfield  {title} {\bibinfo {title} {Thermal conductivity of
  anisotropic and frustrated spin-$\frac{1}{2}$ chains},\ }\href
  {https://doi.org/10.1103/PhysRevB.66.140406} {\bibfield  {journal} {\bibinfo
  {journal} {Phys. Rev. B}\ }\textbf {\bibinfo {volume} {66}},\ \bibinfo
  {pages} {140406} (\bibinfo {year} {2002})}\BibitemShut {NoStop}%
\bibitem [{\citenamefont {Saito}(2003)}]{PhysRevB.67.064410}%
  \BibitemOpen
  \bibfield  {author} {\bibinfo {author} {\bibfnamefont {K.}~\bibnamefont
  {Saito}},\ }\bibfield  {title} {\bibinfo {title} {Transport anomaly in the
  low-energy regime of spin chains},\ }\href
  {https://doi.org/10.1103/PhysRevB.67.064410} {\bibfield  {journal} {\bibinfo
  {journal} {Phys. Rev. B}\ }\textbf {\bibinfo {volume} {67}},\ \bibinfo
  {pages} {064410} (\bibinfo {year} {2003})}\BibitemShut {NoStop}%
\bibitem [{\citenamefont {Orignac}\ \emph {et~al.}(2003)\citenamefont
  {Orignac}, \citenamefont {Chitra},\ and\ \citenamefont
  {Citro}}]{PhysRevB.67.134426}%
  \BibitemOpen
  \bibfield  {author} {\bibinfo {author} {\bibfnamefont {E.}~\bibnamefont
  {Orignac}}, \bibinfo {author} {\bibfnamefont {R.}~\bibnamefont {Chitra}},\
  and\ \bibinfo {author} {\bibfnamefont {R.}~\bibnamefont {Citro}},\ }\bibfield
   {title} {\bibinfo {title} {Thermal transport in one-dimensional spin gap
  systems},\ }\href {https://doi.org/10.1103/PhysRevB.67.134426} {\bibfield
  {journal} {\bibinfo  {journal} {Phys. Rev. B}\ }\textbf {\bibinfo {volume}
  {67}},\ \bibinfo {pages} {134426} (\bibinfo {year} {2003})}\BibitemShut
  {NoStop}%
\bibitem [{\citenamefont {Sakai}\ and\ \citenamefont
  {Klümper}(2003)}]{Sakai_2003}%
  \BibitemOpen
  \bibfield  {author} {\bibinfo {author} {\bibfnamefont {K.}~\bibnamefont
  {Sakai}}\ and\ \bibinfo {author} {\bibfnamefont {A.}~\bibnamefont
  {Klümper}},\ }\bibfield  {title} {\bibinfo {title} {Non-dissipative thermal
  transport in the massive regimes of {theXXZchain}},\ }\href
  {https://doi.org/10.1088/0305-4470/36/46/006} {\bibfield  {journal} {\bibinfo
   {journal} {Journal of Physics A: Mathematical and General}\ }\textbf
  {\bibinfo {volume} {36}},\ \bibinfo {pages} {11617} (\bibinfo {year}
  {2003})}\BibitemShut {NoStop}%
\bibitem [{\citenamefont {Cardy}(2014)}]{cardy14}%
  \BibitemOpen
  \bibfield  {author} {\bibinfo {author} {\bibfnamefont {J.}~\bibnamefont
  {Cardy}},\ }\bibfield  {title} {\bibinfo {title} {Thermalization and revivals
  after a quantum quench in conformal field theory},\ }\href
  {https://doi.org/10.1103/PhysRevLett.112.220401} {\bibfield  {journal}
  {\bibinfo  {journal} {Phys. Rev. Lett.}\ }\textbf {\bibinfo {volume} {112}},\
  \bibinfo {pages} {220401} (\bibinfo {year} {2014})}\BibitemShut {NoStop}%
\bibitem [{\citenamefont {Schultz}\ \emph {et~al.}(1964)\citenamefont
  {Schultz}, \citenamefont {Mattis},\ and\ \citenamefont {Lieb}}]{schultz64}%
  \BibitemOpen
  \bibfield  {author} {\bibinfo {author} {\bibfnamefont {T.~D.}\ \bibnamefont
  {Schultz}}, \bibinfo {author} {\bibfnamefont {D.~C.}\ \bibnamefont
  {Mattis}},\ and\ \bibinfo {author} {\bibfnamefont {E.~H.}\ \bibnamefont
  {Lieb}},\ }\bibfield  {title} {\bibinfo {title} {Two-dimensional ising model
  as a soluble problem of many fermions},\ }\href
  {https://doi.org/10.1103/RevModPhys.36.856} {\bibfield  {journal} {\bibinfo
  {journal} {Rev. Mod. Phys.}\ }\textbf {\bibinfo {volume} {36}},\ \bibinfo
  {pages} {856} (\bibinfo {year} {1964})}\BibitemShut {NoStop}%
\bibitem [{\citenamefont {Di~Francesco}\ \emph {et~al.}(1997)\citenamefont
  {Di~Francesco}, \citenamefont {Mathieu},\ and\ \citenamefont
  {Sénéchal}}]{francesco97}%
  \BibitemOpen
  \bibfield  {author} {\bibinfo {author} {\bibfnamefont {P.}~\bibnamefont
  {Di~Francesco}}, \bibinfo {author} {\bibfnamefont {P.}~\bibnamefont
  {Mathieu}},\ and\ \bibinfo {author} {\bibfnamefont {D.}~\bibnamefont
  {Sénéchal}},\ }\href {https://doi.org/10.1007/978-1-4612-2256-9} {\emph
  {\bibinfo {title} {{Conformal field theory}}}},\ Graduate texts in
  contemporary physics\ (\bibinfo  {publisher} {Springer},\ \bibinfo {address}
  {New York, NY},\ \bibinfo {year} {1997})\BibitemShut {NoStop}%
\bibitem [{\citenamefont {Landau}\ and\ \citenamefont
  {Lifshitz}(1987)}]{Landau1987Fluid}%
  \BibitemOpen
  \bibfield  {author} {\bibinfo {author} {\bibfnamefont {L.~D.}\ \bibnamefont
  {Landau}}\ and\ \bibinfo {author} {\bibfnamefont {E.~M.}\ \bibnamefont
  {Lifshitz}},\ }\href@noop {} {\emph {\bibinfo {title} {Fluid Mechanics}}},\
  \bibinfo {edition} {2nd}\ ed.\ (\bibinfo  {publisher}
  {Butterworth-Heinemann},\ \bibinfo {address} {Oxford, England},\ \bibinfo
  {year} {1987})\BibitemShut {NoStop}%
\bibitem [{\citenamefont {{Oshikawa}}\ and\ \citenamefont
  {{Watanabe}}(2019)}]{2019arXiv190701212O}%
  \BibitemOpen
  \bibfield  {author} {\bibinfo {author} {\bibfnamefont {M.}~\bibnamefont
  {{Oshikawa}}}\ and\ \bibinfo {author} {\bibfnamefont {H.}~\bibnamefont
  {{Watanabe}}},\ }\bibfield  {title} {\bibinfo {title} {{Quantum Quench and
  $f$-Sum Rules on Linear and Non-linear Conductivities}},\ }\href@noop {}
  {\bibfield  {journal} {\bibinfo  {journal} {arXiv e-prints}\ ,\ \bibinfo
  {eid} {arXiv:1907.01212}} (\bibinfo {year} {2019})},\ \Eprint
  {https://arxiv.org/abs/1907.01212} {arXiv:1907.01212 [cond-mat.str-el]}
  \BibitemShut {NoStop}%
\bibitem [{\citenamefont {{Watanabe}}\ \emph {et~al.}(2020)\citenamefont
  {{Watanabe}}, \citenamefont {{Liu}},\ and\ \citenamefont
  {{Oshikawa}}}]{2020JSP...tmp..238W}%
  \BibitemOpen
  \bibfield  {author} {\bibinfo {author} {\bibfnamefont {H.}~\bibnamefont
  {{Watanabe}}}, \bibinfo {author} {\bibfnamefont {Y.}~\bibnamefont {{Liu}}},\
  and\ \bibinfo {author} {\bibfnamefont {M.}~\bibnamefont {{Oshikawa}}},\
  }\bibfield  {title} {\bibinfo {title} {{On the General Properties of
  Non-linear Optical Conductivities}},\ }\bibfield  {journal} {\bibinfo
  {journal} {Journal of Statistical Physics}\ }\href
  {https://doi.org/10.1007/s10955-020-02654-5} {10.1007/s10955-020-02654-5}
  (\bibinfo {year} {2020}),\ \Eprint {https://arxiv.org/abs/2004.04561}
  {arXiv:2004.04561 [cond-mat.stat-mech]} \BibitemShut {NoStop}%
\bibitem [{\citenamefont {Tanikawa}\ \emph {et~al.}(2021)\citenamefont
  {Tanikawa}, \citenamefont {Takasan},\ and\ \citenamefont
  {Katsura}}]{PhysRevB.103.L201120}%
  \BibitemOpen
  \bibfield  {author} {\bibinfo {author} {\bibfnamefont {Y.}~\bibnamefont
  {Tanikawa}}, \bibinfo {author} {\bibfnamefont {K.}~\bibnamefont {Takasan}},\
  and\ \bibinfo {author} {\bibfnamefont {H.}~\bibnamefont {Katsura}},\
  }\bibfield  {title} {\bibinfo {title} {Exact results for nonlinear drude
  weights in the spin-$\frac{1}{2}$ xxz chain},\ }\href
  {https://doi.org/10.1103/PhysRevB.103.L201120} {\bibfield  {journal}
  {\bibinfo  {journal} {Phys. Rev. B}\ }\textbf {\bibinfo {volume} {103}},\
  \bibinfo {pages} {L201120} (\bibinfo {year} {2021})}\BibitemShut {NoStop}%
\bibitem [{\citenamefont {Tanikawa}\ and\ \citenamefont
  {Katsura}(2021)}]{tanikawa2021fine}%
  \BibitemOpen
  \bibfield  {author} {\bibinfo {author} {\bibfnamefont {Y.}~\bibnamefont
  {Tanikawa}}\ and\ \bibinfo {author} {\bibfnamefont {H.}~\bibnamefont
  {Katsura}},\ }\href@noop {} {\bibinfo {title} {Fine structure of the
  nonlinear drude weights in the spin-1/2 xxz chain}} (\bibinfo {year}
  {2021}),\ \Eprint {https://arxiv.org/abs/2107.13784} {arXiv:2107.13784
  [cond-mat.str-el]} \BibitemShut {NoStop}%
\bibitem [{\citenamefont {{Takasan}}\ \emph {et~al.}()\citenamefont
  {{Takasan}}, \citenamefont {{Tanikawa}},\ and\ \citenamefont
  {{Katsura}}}]{katsura}%
  \BibitemOpen
  \bibfield  {author} {\bibinfo {author} {\bibfnamefont {K.}~\bibnamefont
  {{Takasan}}}, \bibinfo {author} {\bibfnamefont {Y.}~\bibnamefont
  {{Tanikawa}}},\ and\ \bibinfo {author} {\bibfnamefont {H.}~\bibnamefont
  {{Katsura}}},\ }\bibinfo {note} {in preparation}\BibitemShut {NoStop}%
\bibitem [{\citenamefont {{Kapustin}}\ and\ \citenamefont
  {{Spodyneiko}}(2020)}]{2020PhRvB.101d5137K}%
  \BibitemOpen
  \bibfield  {author} {\bibinfo {author} {\bibfnamefont {A.}~\bibnamefont
  {{Kapustin}}}\ and\ \bibinfo {author} {\bibfnamefont {L.}~\bibnamefont
  {{Spodyneiko}}},\ }\bibfield  {title} {\bibinfo {title} {{Thermal Hall
  conductance and a relative topological invariant of gapped two-dimensional
  systems}},\ }\href {https://doi.org/10.1103/PhysRevB.101.045137} {\bibfield
  {journal} {\bibinfo  {journal} {\prb}\ }\textbf {\bibinfo {volume} {101}},\
  \bibinfo {eid} {045137} (\bibinfo {year} {2020})},\ \Eprint
  {https://arxiv.org/abs/1905.06488} {arXiv:1905.06488 [cond-mat.str-el]}
  \BibitemShut {NoStop}%
\bibitem [{\citenamefont {Suzuki}(1976)}]{10.1143/PTP.56.1454}%
  \BibitemOpen
  \bibfield  {author} {\bibinfo {author} {\bibfnamefont {M.}~\bibnamefont
  {Suzuki}},\ }\bibfield  {title} {\bibinfo {title} {{Relationship between
  d-Dimensional Quantal Spin Systems and (d+1)-Dimensional Ising Systems:
  Equivalence, Critical Exponents and Systematic Approximants of the Partition
  Function and Spin Correlations}},\ }\href
  {https://doi.org/10.1143/PTP.56.1454} {\bibfield  {journal} {\bibinfo
  {journal} {Progress of Theoretical Physics}\ }\textbf {\bibinfo {volume}
  {56}},\ \bibinfo {pages} {1454} (\bibinfo {year} {1976})},\ \Eprint
  {https://arxiv.org/abs/https://academic.oup.com/ptp/article-pdf/56/5/1454/5264429/56-5-1454.pdf}
  {https://academic.oup.com/ptp/article-pdf/56/5/1454/5264429/56-5-1454.pdf}
  \BibitemShut {NoStop}%
\bibitem [{\citenamefont {Suzuki}(1985)}]{PhysRevB.31.2957}%
  \BibitemOpen
  \bibfield  {author} {\bibinfo {author} {\bibfnamefont {M.}~\bibnamefont
  {Suzuki}},\ }\bibfield  {title} {\bibinfo {title} {Transfer-matrix method and
  monte carlo simulation in quantum spin systems},\ }\href
  {https://doi.org/10.1103/PhysRevB.31.2957} {\bibfield  {journal} {\bibinfo
  {journal} {Phys. Rev. B}\ }\textbf {\bibinfo {volume} {31}},\ \bibinfo
  {pages} {2957} (\bibinfo {year} {1985})}\BibitemShut {NoStop}%
\bibitem [{\citenamefont {Suzuki}\ and\ \citenamefont
  {Inoue}(1987)}]{10.1143/PTP.78.787}%
  \BibitemOpen
  \bibfield  {author} {\bibinfo {author} {\bibfnamefont {M.}~\bibnamefont
  {Suzuki}}\ and\ \bibinfo {author} {\bibfnamefont {M.}~\bibnamefont {Inoue}},\
  }\bibfield  {title} {\bibinfo {title} {{The ST-Transformation Approach to
  Analytic Solutions of Quantum Systems. I: General Formulations and Basic
  Limit Theorems}},\ }\href {https://doi.org/10.1143/PTP.78.787} {\bibfield
  {journal} {\bibinfo  {journal} {Progress of Theoretical Physics}\ }\textbf
  {\bibinfo {volume} {78}},\ \bibinfo {pages} {787} (\bibinfo {year} {1987})},\
  \Eprint
  {https://arxiv.org/abs/https://academic.oup.com/ptp/article-pdf/78/4/787/5275006/78-4-787.pdf}
  {https://academic.oup.com/ptp/article-pdf/78/4/787/5275006/78-4-787.pdf}
  \BibitemShut {NoStop}%
\bibitem [{\citenamefont {Pirvu}\ \emph {et~al.}(2010)\citenamefont {Pirvu},
  \citenamefont {Murg}, \citenamefont {Cirac},\ and\ \citenamefont
  {Verstraete}}]{pirvu10}%
  \BibitemOpen
  \bibfield  {author} {\bibinfo {author} {\bibfnamefont {B.}~\bibnamefont
  {Pirvu}}, \bibinfo {author} {\bibfnamefont {V.}~\bibnamefont {Murg}},
  \bibinfo {author} {\bibfnamefont {J.~I.}\ \bibnamefont {Cirac}},\ and\
  \bibinfo {author} {\bibfnamefont {F.}~\bibnamefont {Verstraete}},\ }\bibfield
   {title} {\bibinfo {title} {Matrix product operator representations},\
  }\href@noop {} {\bibfield  {journal} {\bibinfo  {journal} {New J. Phys.}\
  }\textbf {\bibinfo {volume} {12}},\ \bibinfo {pages} {025012} (\bibinfo
  {year} {2010})}\BibitemShut {NoStop}%
\bibitem [{\citenamefont {Rams}\ \emph {et~al.}(2015)\citenamefont {Rams},
  \citenamefont {Zauner}, \citenamefont {Bal}, \citenamefont {Haegeman},\ and\
  \citenamefont {Verstraete}}]{rams15}%
  \BibitemOpen
  \bibfield  {author} {\bibinfo {author} {\bibfnamefont {M.~M.}\ \bibnamefont
  {Rams}}, \bibinfo {author} {\bibfnamefont {V.}~\bibnamefont {Zauner}},
  \bibinfo {author} {\bibfnamefont {M.}~\bibnamefont {Bal}}, \bibinfo {author}
  {\bibfnamefont {J.}~\bibnamefont {Haegeman}},\ and\ \bibinfo {author}
  {\bibfnamefont {F.}~\bibnamefont {Verstraete}},\ }\bibfield  {title}
  {\bibinfo {title} {Truncating an exact matrix product state for the xy model:
  Transfer matrix and its renormalization},\ }\href
  {https://doi.org/10.1103/PhysRevB.92.235150} {\bibfield  {journal} {\bibinfo
  {journal} {Phys. Rev. B}\ }\textbf {\bibinfo {volume} {92}},\ \bibinfo
  {pages} {235150} (\bibinfo {year} {2015})}\BibitemShut {NoStop}%
\bibitem [{\citenamefont {Yang}\ \emph {et~al.}(2009)\citenamefont {Yang},
  \citenamefont {Wang}, \citenamefont {Xu}, \citenamefont {Qin},\ and\
  \citenamefont {Xiang}}]{Yang_2009}%
  \BibitemOpen
  \bibfield  {author} {\bibinfo {author} {\bibfnamefont {L.~P.}\ \bibnamefont
  {Yang}}, \bibinfo {author} {\bibfnamefont {Y.~J.}\ \bibnamefont {Wang}},
  \bibinfo {author} {\bibfnamefont {W.~H.}\ \bibnamefont {Xu}}, \bibinfo
  {author} {\bibfnamefont {M.~P.}\ \bibnamefont {Qin}},\ and\ \bibinfo {author}
  {\bibfnamefont {T.}~\bibnamefont {Xiang}},\ }\bibfield  {title} {\bibinfo
  {title} {A quantum transfer matrix method for one-dimensional disordered
  electronic systems},\ }\href {https://doi.org/10.1088/0953-8984/21/14/145407}
  {\bibfield  {journal} {\bibinfo  {journal} {Journal of Physics: Condensed
  Matter}\ }\textbf {\bibinfo {volume} {21}},\ \bibinfo {pages} {145407}
  (\bibinfo {year} {2009})}\BibitemShut {NoStop}%
\end{thebibliography}%

\end{document}